\documentclass[fleqn,usenatbib]{mnras}
\usepackage[T1]{fontenc}
\usepackage{ae,aecompl}
\usepackage{graphicx}   % Including figure files
\usepackage{amsmath}    % Advanced maths commands
\usepackage{amssymb}    % Extra maths symbols
\usepackage{float}
\usepackage[normalem]{ulem}
\usepackage[shortlabels]{enumitem}

\usepackage[dvipsnames]{xcolor}

\title[Stratified jet synchrotron emission]%
{Synchrotron intensity plots from a relativistic stratified jet}
\author[Frolova et al.]{\parbox{\textwidth}{
V.~A.~Frolova$^{1}$\thanks{E-mail: frolova.va@phystech.edu},
E.~E.~Nokhrina$^{1}$, 
I.~N.~Pashchenko$^{2}$
}
\vspace{0.4cm}\\
\parbox{\textwidth}{
$^1$Moscow Institute of Physics and Technology, Dolgoprudny, Institutsky per., 9, Moscow region, 141700, Russia\\
$^2$Lebedev Physical Institute, Leninsky prosp.~53, Moscow, 119991, Russia
}
}

\begin{document}

\date{Accepted 2023 May 4. Received 2023 May 4; in original form 2023 March 13}

\pagerange{\pageref{firstpage}--\pageref{lastpage}} \pubyear{}

\maketitle

\label{firstpage}

\begin{abstract}

We examine the effect of a jet transversal structure from magnetohydrodynamic semi-analytical modelling on the total intensity profiles of relativistic jets from active galactic nuclei. In order to determine the conditions for forming double- and triple-peaked transverse intensity profiles, we calculate the radiative transfer for synchrotron emission with self-absorption from the jets described by the models with a constant angular velocity and with a total electric current closed inside a jet. We show that double-peaked profiles appear either in the models with high maximal Lorentz factors or in optically thick conditions. We show that triple-peaked profiles in radio galaxies constrain the fraction of the emitting particles in a jet. We introduce the possible conditions for triple-peaked profiles under the assumptions that nonthermal electrons are preferably located at the jet edges or are distributed according to Ohmic heating.
\end{abstract}


\section{Introduction}
\label{s:intro}

Relativistic jets from active galactic nuclei (AGN) are thought to originate in the vicinity of supermassive black holes (SMBH) having angular momentum (spin) and accreting matter from the ambient medium \citep{BMR-19}. Relating the observations of jets on parsec scales in radio and millimeter band to the physics underlying AGN activity and jet production requires more or less elaborate modelling of jet and emitting plasma properties. The Blandford--K\"onigl model \citep{BlandfordKoenigl1979} employs the assumption of equipartition of emitting plasma energy density with the proper magnetic field energy density \citep{Burbidge56}. The model assumes the certain dependencies of particle number density and magnetic field on the distance from the jet base. Despite neglecting the jet transversal stratification, this model was successful in explaining such observational phenomena as core shift effect \citep{Kovalev08}. The Blandford--K\"onigl model was used to estimate the jet magnetic field and plasma number density \citep{L98, hirotani2005, Sullivan09_coreshift, NBKZ15}, although for extreme brightness temperatures a proper accounting for the non-uniform jet structure is needed \citep{Nok17A}. At the same time, theoretical and numerical modelling of non-uniform jets within the approach of ideal magnetohydrodynamics (MHD) progressed to reproducing plasma bulk acceleration up to relativistic velocities \citep{Beskin06, Kom09, Lyu09, TMN09} and a jet boundary shape \citep{Chatterjee2019, Nokhrina19, Kov-20}. 

The recent jet observations by the very-long-baseline interferometry (VLBI) method in radio band allow us to look at the jets down to the event horizon scales \citep{EHT_I, EHT_V}. These observations reveal the clear jet stratification in intensity with a pronounced limb brightening in Mrk501 \citep{Giroletti04}, M87 \citep{Walker18}, 3C 84 \citep{Giov18} and Cen A \citep{Janssen21}; transition from spine to limb brightening at different frequencies in 3C 273 \citep{Bruni2021} and even a triple-peaked emission structure in the core region in M87 \citep{Hada17} which is detected down to the scales of a few gravitational radii \citep{Lu23}. 

There may be several reasons for the observed stratified emission from relativistic jets. The stratified poloidal velocity field may lead to boosting and de-boosting of emission from different parts of a jet. 
In particular, in the models with fast spine and slow sheath, limb brightening may occur due to de-boosting of emission from a sheath, while the emission from a fast relativistic spine is concentrated in the narrow angle around the bulk plasma velocity, and we do not observe this emission. Another possible source of a stratified emission is a spatial distribution of nonthermal electrons. We assume that the synchrotron self-absorbed emission is produced by the relativistic plasma with energy distribution given by the power law 
\begin{equation}
{\rm d}n_{\rm emit}=k_e(\textbf{r})\gamma^{-p}{\rm d}\gamma,
\label{nofE}
\end{equation} 
where $\gamma$ is a particle Lorentz factor, $n_{\rm emit}$ is emitting plasma number density with the amplitude $k_e$ depending on the coordinate $\textbf{r}$ within a jet, $p$ is a power law index related to the spectral index $\alpha = (1 - p) / 2$ of the synchrotron emission optically thin regime spectral flux 
\begin{equation}
S_{\nu}\propto \nu^{\alpha}.
\end{equation}
Such distribution may be produced by the acceleration on shocks \citep[e.g.,][]{Kirk1994} by the Fermi mechanism of the first order or by the magnetic reconnection \citep{SironiSpitkovskyArons2013} with subsequent appearance of region with an electric field larger than a magnetic one, where charged particles accelerate effectively. These processes may take place locally, for example, at the jet boundary due to developing instabilities (e.g. pinch instabilities shown by \citet{McKinney2006, Chatterjee2019}, Kelvin-Helmholtz instability \citep[e.g.][Nikonov et al. submitted]{Hardee2011}). The power-law spectrum can also be produced by shear acceleration when the particles gain energy crossing the tangential velocity discontinuity at the jet boundary due to diffusion, if the velocity jump is relativistic and the turbulence is present \citep{Ostrowski1990, Ostrowski1998}. So, nonthermal electrons may also be distributed non-uniformly across a jet, giving rise to 
peculiar structures in intensity and spectral maps 
\citep[e.g. Nikonov et al. submitted,][]{Bruni2021}.
Non-uniform distribution of nonthermal electrons could be also due to possible particle acceleration as a result of Ohmic heating, so that the amplitude $k_e$ is proportional to the square of co-moving electric current density $j'^2$ \citep{Lyutikov2005}, or due to the equipartition regime between plasma and magnetic field, so that $k_e$ is proportional to the square of proper magnetic field $B'^2$ \citep{Burbidge56, BlandfordKoenigl1979}. One more reason may be a non-uniform jet structure itself: as the emission and absorption coefficients depend on a local magnetic field and particle number density, variation in these physical values leads to variations in emission. This approach has been used to estimate the magnetic field in a jet by the observed brightness temperature \citep{Nok17A}. Another possible reason of the observed stratification is the opacity effect. \cite{2008ApJ...679..990Z} considered a ``core-dominated'' jet with optically thick geometrically thin core and a surrounding optically thin jet. It was shown that
the observed transverse structure could significantly depend on frequency. Observations of quasar 3C~273 with RadioAstron at 1.6 and 4.8 GHz by \cite{Bruni2021} revealed the effect of a frequency-dependent intensity profile. The jet from 3C~273 is limb-brightened at 1.6 GHz and is spine-brightened at 4.8 GHz, with a clear spectral index gradient across the jet. Authors considered several concurring factors that could explain the observed picture, including plasma stratification across the jet. There is also a possibility that many of these effects contribute to the observed emission profiles \citep[see, e.g.,][]{Gabuzda21}.

To model observed intensity profiles from M87 jet, steady axisymmetric force-free jet models have been successfully employed. \citet{BroderickLoeb2009} used the Gaussian distributions of both thermal and nonthermal electrons and showed that limb brightened intensity profiles emerge in the case of high black hole (BH) spin, and low spin produces one-peaked, highly asymmetric profiles. Setting the Gaussian distribution of emitting electrons at jet's base in the form of a ring, \citet{Takahashi2018} showed that the symmetrical intensity profiles require a fast-spinning BH as well, and \citet{Ogihara19} reproduced the triple-peaked intensity profile for the optically thin jet. Relativistic and general relativistic magnetohydrodynamical (RMHD and GRMHD) simulations are also used for modelling M87 intensity profiles. Within GRMHD simulations, \citet{MFS16} obtained a limb-brightened intensity profile using the Monte-Carlo technique developed for relativistic radiative transfer. RMHD simulations by \citet{Fuentesetal2018} show the three-peaked profiles for the models with high magnetization.
\citet{KramerMacDonald2021} explored different magnetic field morphologies and electron distributions using 3D RMHD simulations and found out that the jets with purely toroidal and poloidal magnetic fields are limb-brightened and spine-brightened correspondingly, while the electron distributions set proportional to thermal, internal and magnetic energy densities produce similar results. The equipartition between the magnetic and thermal pressure was assumed. Though the intensity profiles patterns are commonly reproduced, the realistic, self-consistent magnetic fields and electron number densities are not well studied. 

Analytical and semi-analytical magnetohydrodynamical (MHD) models \citep{BesMal-2000, Beskin06, BCKN-17} provide stationary distributions of physical parameters: velocity, magnetic field $\textbf{B}$, particle number density $n$. We should emphasize that $n$ obtained within this approach is a total (cold) particle number density. The emission and absorption coefficients depend on the nonthermal emitting particle number density $n_\mathrm{emit}\leqslant n$ with the assumed power law energy distribution (\ref{nofE}). The particular value of $n_\mathrm{emit}$ depends on the model assumptions. However, all the physical quantities obtained within the axisymmetrical stationary MHD approach do not depend on time, have smooth distributions and, thus, can be easily used in the numerical calculation of the radiative transfer equation. We use two models described by \citet{Beskin06, Lyu09} and by \citet{BCKN-17}. Both models predict the jet properties in very good agreement with the numerical modelling by \citet{Kom07} and by \citet{Chatterjee2019}.

Using the modelled emission profiles, we explore their dependencies on such parameters as a viewing angle, separation from the central engine, total magnetic flux and light cylinder radius. 
We examine the contributions to the intensity profiles made by the Doppler factor, jet stratification and different distributions of relativistic plasma across a jet. Our objective is to explain the most probable origins of peculiarities observed in jets and to constrain the possible models for nonthermal electrons spatial distribution.

The paper is organized as follows. In \autoref{s:comp} we discuss the chosen models and explain the key principles used to carry out the simulations. Then, we present achieved results in \autoref{s:intensity} focusing on the explanation of what profiles should appear with the use of the models in consideration and on the intensity profiles dependencies on the models parameters. In \autoref{s:heating}, the possible model modifications for limb brightening production are discussed. In \autoref{s:discussion}, we compare our results to the works by other authors and specifically to Blandford-K{\"o}nigl model. Finally, we summarize our work in \autoref{s:summary}.

\section{Approach}
\label{s:comp}

\subsection{Two models of a jet transversal structure}
\label{ss:model}

To model jet synchrotron emission, we use the resultant jet structure obtained within the approach of stationary axisymmetric magnetohydrodynamics. The distribution of magnetic field $\textbf{B}$, particle number density $n$ and bulk Lorentz factor $\Gamma$ are governed by Grad--Shafranov and Bernoulli equations. We assume the electron--positron jet composition \citep{Zdziarski-22a, Zdziarski-22b}. The  solution depends on the particular choice of five quantities, conserved on magnetic surfaces $\Psi=\mathrm{const}$, where $\Psi$ is magnetic flux, --- integrals of energy density flux $E(\Psi)$, angular momentum density flux $L(\Psi)$, angular velocity $\Omega_{\textbf{F}}(\Psi)$, particle density flux to magnetic field ratio $\eta(\Psi)$ and entropy $s(\Psi)$ \citep[see details in][]{Beskin10}. After the choice of integrals and for the highly collimated flows we can use the cylindrical approach, so the Grad--Shafranov and Bernoulli equations reduce to the set of ordinary differential equations. The particular solution depends on the integrals choice, and we employ two models. The first one (M1) describes a jet with a constant angular velocity $\Omega_\mathrm{F}(\Psi)=\Omega_0$ and linear energy and angular momentum integrals \citep[see the particular integrals choice in][]{Beskin06, Lyu09}. The distribution of physical quantities across a jet are presented in Figures~3--4 by \citet{NBKZ15} and are in good agreement with the numerical modelling by \citet{Kom07, BTch-16}.

As the second model (M2) we use the model by \citet{BCKN-17}. It consists of a central part of a flow alike the Model~1. But due to the special choice of the integrals 
\begin{equation}
\Omega_\mathrm{F}(\Psi)=\Omega_0\sqrt{1-\Psi/\Psi_0},
\label{OmegaM2}
\end{equation}
$E(\Psi)$ and $L(\Psi)$, the model has a slower sheath --- a mildly to non-relativistic outflow around a fast central part. The distributions of physical quantities are presented in Figures~1--5 in \citet{CBP19}. In both models the jet boundary is set as a radius, at which the magnetic flux becomes equal to the total magnetic flux $\Psi_0$ contained in a jet.

These models describe identically the central part of a jet, where the field lines angular velocity $\Omega_{\rm F}$ is almost constant. Below $r$ denotes the cylindrical radius.

{\it 1. Central core, $r<$ a few $R_\mathrm{L}$.}

The central dense core is a part of a jet with the radius of several light cylinder radii $R_{\rm L}=c/\Omega_{\rm F}$. It characterizes by almost constant particle number density, which is highest across a jet, and a uniform poloidal magnetic field \citep{Kom07, Beskin09, Lyu09}. Here $c$ is the speed of light. The toroidal magnetic field in the central core is less than or comparable to the poloidal magnetic field, and the flow bulk Lorentz factor is equal to the initial Lorentz factor$\Gamma_{\rm in}$, as this part of a jet remains plasma-dominated from the base. Presence of such a dense jet core is a robust feature of both analytical and numerical models, and it is important to understand its impact on the overall jet emission and possible limitations on the location of nonthermal emitting electrons. 

{\it 2. Intermediate part, $r<$ a few $R_\mathrm{L}$ and $\Psi<\Psi_0/2$.}

Outside this central core and for the field lines with constant angular velocity, which is present in both M1 and M2 (up to $\Psi\approx\Psi_0/2$), poloidal $B_{\rm P}$ and toroidal $B_{\varphi}$ magnetic fields and a particle number density $n$ decrease with the radial distance while the flow accelerates. This leads to a drop in emissivity and in plasma absorption, and to the strong boosting of emission within a small cone with half-opening angle $\sim\Gamma^{-1}$ around the bulk motion velocity direction.

{\it 3. Outer part, $\Psi>\Psi_0/2$.}

Within M1, the acceleration continues up to the jet boundary with $\Psi=\Psi_0$, where the highest Lorentz factor is achieved by the flow. The values of $n$ and $B$ are the decreasing functions of $r$. In contrast, in the M2 the flow bulk velocity decreases towards the jet boundary, leading to emission de-boosting. Particle number density rises abruptly at the boundary \citep[see Figure~1 by][]{CBP19} as it is the thermal pressure that balances the ambient medium pressure in the M2. So we expect to explore the impact of slower velocity in producing the limb-brightening effect, observed in different sources. The same qualitative behaviour of a bulk Lorentz factor, as in the M2, has been obtained in the numerical simulations by \citet{Chatterjee2019}: the flow becomes slower towards the jet boundary, and the Lorentz factor peaks closer to the jet axis relatively to the jet width as compared with further jet cross-sections.

We use the following prescription on the emitting plasma particle number density $n_\mathrm{emit}$. In Section~\ref{s:intensity} we assume that all the plasma from MHD modelling has a power-law distribution (\ref{nofE}) and emits, thus $n_\mathrm{emit}=n$. This is done in order to capture the impact of the jet stratification on the total intensity profiles. In Section~\ref{s:heating} we set the emitting plasma number density proportional to the magnetic field energy density \citep{Burbidge56, BlandfordKoenigl1979, L98} and to the electric current density in the proper plasma frame \citep{Lyutikov2005}.

\subsection{Calculating the emission profiles}
\label{ss:comp}

The problem is considered within the cylindrical approach: we model different intensity profiles across a jet assuming it to be a cylinder with no physical quantities evolution along a jet. This approach works well for the optically thick regimes, as the jet longitudinal evolution is very slow. For optically thin parts of a jet observed at large enough observational angles, the cylindrical approach also is a good approximation, as the profile is formed from the emission of nearby jet slices. In a case of an optically thin jet observed at small angle such approach cannot be applied to modelling the particular jet emission profile, but is aimed to catch the robust features of an emission independently of any model for a longitudinal jet evolution. 

To compute the transverse intensity profile, the radiative transfer equation for the synchrotron emission with self-absorption is numerically solved. The radiative transfer equation in the observer frame has the form
\begin{equation}
\label{rte}
    \frac{{\rm d} I_{\nu}}{{\rm d} s} = j_{\nu} - \text{\ae}_{\nu}I_{\nu},
\end{equation}
where $I_{\nu}$ is the spectral intensity, $j_{\nu}$ and \ae$_{\nu}$ are the spectral emission and absorption coefficients, and the equation is solved along the line of sight.

The synchrotron emission and absorption coefficients are written in the plasma proper frame as follows \citep{GinzburgSyrovatskii65}:
\begin{equation}
\label{j}
    j_{\nu'}' =  h(p) \frac{e^3}{mc^2} \left(\frac{3e}{2\pi m^3 c^5}\right)^{(p - 1)/2} \frac{K}{4\pi} B_{\perp}'^{(p + 1)/2} \nu'^{(1 - p)/2},
\end{equation}
\begin{equation}
\label{ae}
    \text{\ae}'_{\nu'} = g(p) \frac{e^2}{2\pi m} \left(\frac{3e}{2\pi m^3 c^5}\right)^{p / 2} K B_{\perp}'^{(p + 2)/2} \nu'^{-(p + 4)/2},
\end{equation}
where
\begin{equation}
    h(p) = \frac{\sqrt{3}}{p + 1}\Gamma\left(\frac{3p - 1}{12}\right) \Gamma\left(\frac{3p + 19}{12}\right),
\end{equation}
\begin{equation}
    g(p) = \frac{\sqrt{3}}{4}\Gamma\left(\frac{3p + 2}{12}\right) \Gamma\left(\frac{3p + 22}{12}\right).
\end{equation}
Here $K = k_e (mc^2)^{p - 1}$, 
$e$ is the electron charge, $m$ is the electron mass, $B_{\perp}'$ is the magnetic field component perpendicular to the wave vector, $\Gamma$ is gamma function. All the primed quantities and $k_e$ are in the plasma proper frame. The emission and absorption coefficients are recalculated to the lab frame using the Lorentz invariants $I_{\nu} / \nu^3$, $j_{\nu} / \nu^2$ and $\text{\ae}_{\nu} \nu$ \citep{RL79}. The proper and observed frequencies $\nu'$ and $\nu$ correspondingly are related via
\begin{equation}
    \frac{\nu'}{\nu} = \frac{1 + z}{\delta},
\end{equation}
where $z$ is cosmological redshift of the source and $\delta$ is Doppler factor. As we intend to describe close, well-resolved sources, we omit $z$.

The distribution \eqref{nofE} is integrated to express the amplitude $k_e$ in terms of the emitting particle number density $n_{\rm emit}$:
\begin{equation}
\label{ke_n}
    k_e = \begin{cases}
        \displaystyle \frac{(p - 1) n_{\rm emit}}{\gamma_{\rm min}^{1 - p} - \gamma_{\rm max}^{1 - p}} \approx (p - 1) \gamma_{\rm min}^{p - 1} n_{\rm emit},\ p \neq 1, \\ \ \\
        \displaystyle \frac{n_{\rm emit}}{\ln \gamma_{\rm max}/\gamma_{\rm min}},\ p = 1.
    \end{cases}
\end{equation}

For short, we further denote
\begin{equation}
    n_{p, \gamma_{\rm min}} =  (p - 1) \gamma_{\rm min}^{p - 1} n_{\rm emit}.
\end{equation}

In the calculations, the minimal Lorentz factor $\gamma_{\rm min}$ is set equal to unity \citep{Wardle98}, and the maximal Lorentz factor $\gamma_{\rm max}$ is neglected as $\gamma_{\rm max}^{1 - p} \ll 1$. The value of $p$ can be determined for a particular source from observations but its theoretical evaluation is under discussion. From a theoretical perspective the numerical particle-in-cell simulations are used to explore the possible values of $p$. For relativistic reconnection, they predict the values between $1.5$ and $4$, with larger values for lower magnetizations \citep{SironiSpitkovsky2014,2018ApJ...862...80B}. On the other hand, for particle acceleration in shocks, they provide $\approx 2.5$ for low magnetizations \citep{SironiSpitkovsky2011}. Hybrid fluid-particle simulations can help to estimate $p$ as well \citep[e.g.][Kramer et al. in prep.]{Vaidya2018}; \citet{Vaidya2018} got the mean of $p \sim 3.1$ in their simulations.
Observationally, \cite{2012A&A...544A..34P} obtained the median spectral index in jets of 319 sources $\alpha = -0.68$ between 2 and 8 GHz. \cite{MOJAVE_XI} found the mean spectral index in the jets of 190 AGNs $\alpha = -1.04\pm0.03$ with more flattened spectrum in jet components using four frequency VLBA observations between 8 and 15 GHz. Recently, \cite{2023arXiv230112861P} employed multifrequency VLBI simulations with relativistic jet model and found that the imaging procedure traditionally used for obtaining the spectral index maps from the VLBI data could artificially steepen the spectrum. Simulations also revealed that the intrinsic optically thin spectral index $\alpha = -0.5$, corresponding to $p = 2$, is consistent with recent dual frequency (8 and 15 GHz) high-sensitivity VLBI observations of M87 radio jet (Nikonov et al. submitted).
In the work, the value $p = 2$ is selected as the basic value for certainty. 

With $\gamma_{\rm min} = 1$ and $p = 2$, the expression \eqref{ke_n} is simply $k_e = n_{\rm emit}$.

\subsection{Setting the model parameters}

Our modelling depends on the following intrinsic parameters. 

Firstly, the MHD modelling provides the cross-sections of the jet. The calculated physical quantities are the functions of the radial coordinate $r$ only, since the equations are solved within the cylindrical approach. For the chosen model, M1 or M2, the cross-section is defined by the initial magnetization $\sigma_{\rm M}$ and dimensionless radius $d_{R_{\rm L}}$ in the units of $R_{\rm L}$. Initial magnetization defines the plasma terminal velocity. Dimensionless jet width is connected with the jet boundary geometry: 
larger jet radii in terms of a light cylinder radius correspond to larger distances along a jet. On the other hand, larger given jet width at the cross-section corresponds to smaller $R_{\rm L}$ or, equivalently, a transverse size of a central core. 

To solve the radiative transfer equation we must use the dimensional variables. For the particle number density $n$ and the magnetic field $B$ the relation is a follows:
\begin{equation}
\label{ndim}
    n = \tilde{n} \left(\frac{\Psi_0}{8\pi^2 R_{\rm L}^2 \sigma_{\rm M}}\right)^2 \frac{1}{mc^2},
\end{equation}
\begin{equation}
\label{Bdim}
    B = \tilde{B} \frac{\Psi_0}{2\pi \sigma_{\rm M} R_{\rm L}^2},
\end{equation}
where the dimensionless quantities are denoted with the tilde. Here the light cylinder radius can be related to the BH spin, and  the combination of $R_{\rm L}$ and $\Psi_0$ can be used to estimate the jet power.

The total magnetic flux $\Psi_0$ defines the optical thickness of the source (the part of jet under consideration) at a given frequency, and characterises the following jet parameters:
\begin{enumerate}[i.]
\itemindent=20pt%
    \item the optical depth $\tau$ of the source at a given frequency;
    \item the part of the cross-section mostly contributing to the intensity -- the position of the $\tau = 1$ surface in the jet; 
    \item the spectral index between the pair of frequencies.
\end{enumerate}
All these quantities are strongly stratified. Due to the central core-defined jet structure, these values possess the extrema in the central part of the jet; the extrema are shifted from the precise jet spine due to the rotation. 

As we fix the jet local physical width, the light cylinder radius and the initial magnetization, we have only the total magnetic flux $\Psi_0$ to vary. We choose the particular values of $\Psi_0$ to fix the maximum of the optical depth $\tau_{\rm max}$ for the given cross-section. We prefer the optical depth since it is the Lorentz invariant and it is the only stratified characteristic whose transverse profile does not change with $\Psi_0$. Moreover, this way of fixing $\Psi_0$ can be applied independently of the model: though the order of $\Psi_0$ corresponding to the particular $\tau_{\rm max}$ values differs strongly depending on the $({\rm M}, d_{R_{\rm L}}, \sigma_{\rm M})$ tuple, the optical depth itself stays the universal physical characteristic of a source. Thus, the profiles are indexed with the maximum optical depth value, and the corresponding total magnetic flux and jet power will be listed in the table. The maximum optical depth is varied over the whole range covering optically thin and optically thick options -- we tried $15$ values from $0.01$ to $500$, and we chose $8$ values from this set to present in the plots in the paper. The total magnetic flux $\Psi_0$ is also related to the jet power, which can be calculated via the Poynting flux power at the jet base \citep{Nokhrina18}:
\begin{equation}
    W_{\rm jet} = \frac{c}{8} \left(\frac{\Psi_0}{\pi R_{\rm L}}\right)^2.
\end{equation}

In our study we focus on understanding of impact of a jet stratification and different models of the spatial distribution of nonthermal plasma on the total intensity profiles. In order to capture general features, we keep the number of the parameters that we vary as small as possible. To address the jet viewing angle, we choose two values of $5^{\circ}$ and $15^{\circ}$. They are typical for BL Lacs/quasars/blazars and radio galaxies without a pronounced counter-jet correspondingly \citep{MOJAVE_IX, MOJAVE_XV, MOJAVE_XVII, Kov-20}. The value $15^{\circ}$ is also preferable as it lies in the interval of estimated viewing angle for M87 between $14^{\circ}$ and $19^{\circ}$ \citep{Nakamura_Meier_14, Mertens16, Walker18, Nakamura+18, Kim2018}, which is especially interesting source with the best resolved transversal and longitudinal structure in intensity. Below we refer to the sources with these two viewing angles as `BL Lacs/quasars/blazars' and `radio galaxies' meaning only the typical viewing angle, and not the real source type.

For initial magnetization $\sigma_{\rm M}$, \citet{Nokhrina2022} showed that $\sigma_{\rm M} \approx 2 \Gamma_{\rm max}$ and $\Gamma_{\rm max}/2$ for M1 and M2 correspondingly. Thus, for M1, considering M87 as an example, $\sigma_{\rm M} = 5; 20$ are selected as reference values according to $\Gamma \sim 3$ at small scales reported by \citet{Mertens16} and $\Gamma \sim 10$ at larger scales reported by \citet{BSM99}. $\sigma_{\rm M} = 50$ is chosen as well to demonstrate the higher velocities effects. For blazars, Lorentz factors are estimated to be up to $~50$ using VLBI kinematics \citep{MOJAVE_XVII}, so the chosen value can be taken into account in the qualitative considerations. For M2, the corresponding values are modified to maintain the upper velocity limit. The precise $\Gamma_{\rm max}$ at the considered jet cross-sections are given in Table \ref{tab:G_max}.

\begin{table}
    \centering
    \begin{tabular}{c|c|c|c}
    \hline
         Model & $\sigma_{\rm M}$ & $d_{R_{\rm L}}$ & $\Gamma_{\rm max}$ \\
    \hline
         M1 & $5$ & $25$ & $4.0$
         \\
         M1 & $5$ & $100$ & $4.4$
         \\
         M1 & $20$ & $25$ & $9.9$ 
         \\
         M1 & $20$ & $100$ & $13.4$ 
         \\
         M1 & $50$ & $25$ & $15.0$ 
         \\
         M1 & $50$ & $100$ & $25.9$ 
         \\
         M2 & $5$ & $25$ & $3.1$
         \\
         M2 & $5$ & $100$ & $3.7$ 
         \\
         M2 & $20$ & $25$ & $6.6$
         \\
         M2 & $20$ & $100$ & $10.4$
         \\
         M2 & $50$ & $25$ & $9.0$
         \\
         M2 & $50$ & $100$ & $18.5$
         \\
    \hline
    \end{tabular}
    \caption{The maximum Lorentz factor $\Gamma_{\rm max}$ on the cross-sections considered in the paper.}
    \label{tab:G_max}
\end{table}

To set $R_{\rm L}$, two characteristic cross-sections are considered: with the central core taking up i. nearly all the jet width; ii. only the small fraction of the jet width. This is done to explore the impact of a central core on the overall intensity map, as it occupies the scales of several $R_{\rm L}$ \citep{Beskin09}. For certainty, for every model under consideration we choose the universal values $d_{R_{\rm L}} = 25; 100$. These values are treated in two different ways: i. the geometrical width is kept, so the central core impact is explored; ii. the light cylinder radius is kept, so the separation from the central engine is explored.

The geometrical jet width is set to $0.5$ pc and the luminosity distance $0.1$ pc/mas is used for the spectral flux calculation. This fiducial choice aims to provide the outcome expected from the considered models for the nearby resolved on the sub-parsec scales AGN jets. For M87, the closest and best resolved source, the luminosity distance is $0.08$ pc/mas and the sub-parsec jet is few mas wide.

This choice is not restrictive. Suppose rescaling the jet width, in pc, as $d' = a d$ and examining the same model cross-section of the jet. The profile contours at the same optical depths, but at $R_{\rm L}$ and $R_{\rm L}' = a R_{\rm L}$ are almost indistinguishable to the central core profoundness, and the primary difference is the intensity values. The intensity can be approximately recalculated from the homogeneity assumption after equating $\tau$ for $R_{\rm L}$ and $R_{\rm L}'$:
\begin{equation}
\label{eq:recalc_width_real}
    I' \approx a^{1/(p + 6)} I.
\end{equation}

Now suppose rescaling the observed jet width, in mas, as $d' = a d$ and examining the same cross-section of the jet. If the intensity profile is divided into the same number of pixels, then each of them is seen at the solid angle $\Omega' = a^2 \Omega$, so
\begin{equation}
\label{eq:recalc_width_obs}
    I' = a^2 I.
\end{equation}

The total magnetic flux $\Psi_0$ is expected to have the order of $10^{33}-10^{34}$ G cm$^2$ \citep{Zamaninasabetal2014}. As we choose the model parameters as described above, 
total magnetic flux $\Psi_0$ is not predetermined, but chosen in order to achieve the particular value of the optical depth in the central part of the jet. Thus, we analyse and compare its values \textit{a posteriori}. 

The jet rotation is responsible for asymmetries in the intensity profiles, and we plot the advancing side of a jet to the left, and the receding part of a jet to the right.

The model parameters are summarized in Table \ref{tab:par}.

\begin{table}
    \centering
    \begin{tabular}{c|c}
    \hline
         Parameter & Tried values\\
         \hline
         Model & \begin{tabular}[x]{@{}c@{}}M1, M2 \\(see  \autoref{ss:model})\end{tabular} \\
         $\sigma_{\rm M}$ & $5$, $20$, $50$ \\
         $\theta$ & $5^{\circ}$, $15^{\circ}$ \\
         \begin{tabular}[x]{@{}c@{}}
             $(d_{R_{\rm L}}, R_{\rm L})$ \\
             (`$R_{\rm L}$ comparison' mode) 
         \end{tabular} & (25, 0.01 pc), (100, 0.025 pc) \\
          \begin{tabular}[x]{@{}c@{}}
             $(d_{R_{\rm L}}, R_{\rm L})$ \\
             (`evolution' mode) 
         \end{tabular} & (25, 0.01 pc), (100, 0.01 pc) \\
         \hline
    \end{tabular}
    \caption{The parameters used for intensity profiles modelling.}
    \label{tab:par}
\end{table}

\section{Spectral intensity transverse profiles} 
\label{s:intensity}

In this section, 
we present the modelling of
spectral intensity profiles on the $(\rm{M}, \sigma_{\rm M}, d_{R_{\rm L}})$ tuples
and discuss the robust features of emission from MHD jets when relativistic plasma has the same energy distribution across a jet. In \autoref{s:heating}, we explore the effects of non-uniform spatial distribution of radiating plasma.

\subsection{`$R_{\rm L}$ comparison' mode}
\label{ss:BHspin_mode}
\begin{figure}
\centering
\includegraphics[width=\columnwidth, angle=0]{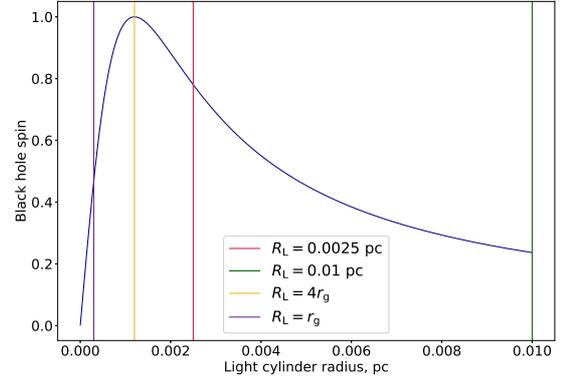}
\caption{The dependence of the absolute value of a BH spin on the light cylinder radius for M87. The vertical lines indicate our probe values, the gravitational radius and the maximum at $4 r_{\rm g}$.}
\label{BHspin_vs_RL}
\end{figure}

\begin{figure*}
    \centering
    \includegraphics[width = \textwidth]{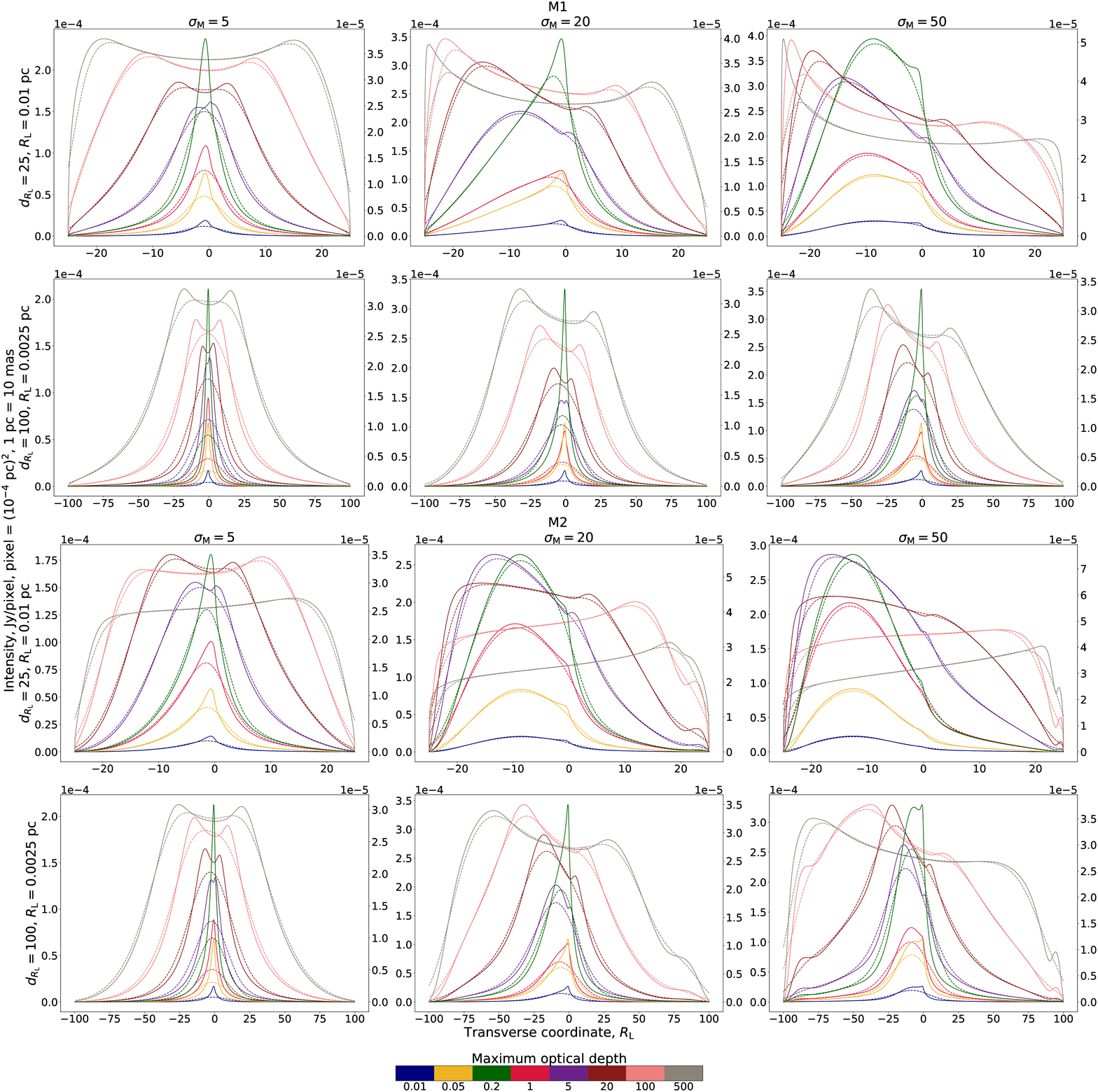}
    \caption{The ideal (solid lines) and convolved (dashed lines) intensity profiles at the viewing angle $5^{\circ}$. The upper and lower plots describe the models M1 and M2 correspondingly (see the models' description in \autoref{ss:model}). The $R_{\rm L}$ values are chosen in the `$R_{\rm L}$ comparison' mode (the first and second row for each model describe the jet cross-section of the same geometrical width, $0.5$ pc, but with different light cylinder radii, see details in the beginning of \autoref{s:intensity}). The colours correspond to the maximum optical depth over the cross-section, $\tau_{\rm max}$. $\tau_{\rm max} \leqslant 1$ are plotted at the left y-axis, and $\tau_{\rm max} < 1$ are plotted at the right y-axis. The convolving beam is round with the FWHM equal to $0.1$ of the cross-section width ($0.5$ mas). The physical parameters for this figure are summarized in Tables \ref{tab:blazars_spin} and \ref{tab_app:blazars_spin}.}
    \label{fig:blazars_spin}
\end{figure*}

\begin{figure*}
    \centering
    \includegraphics[width = \textwidth]{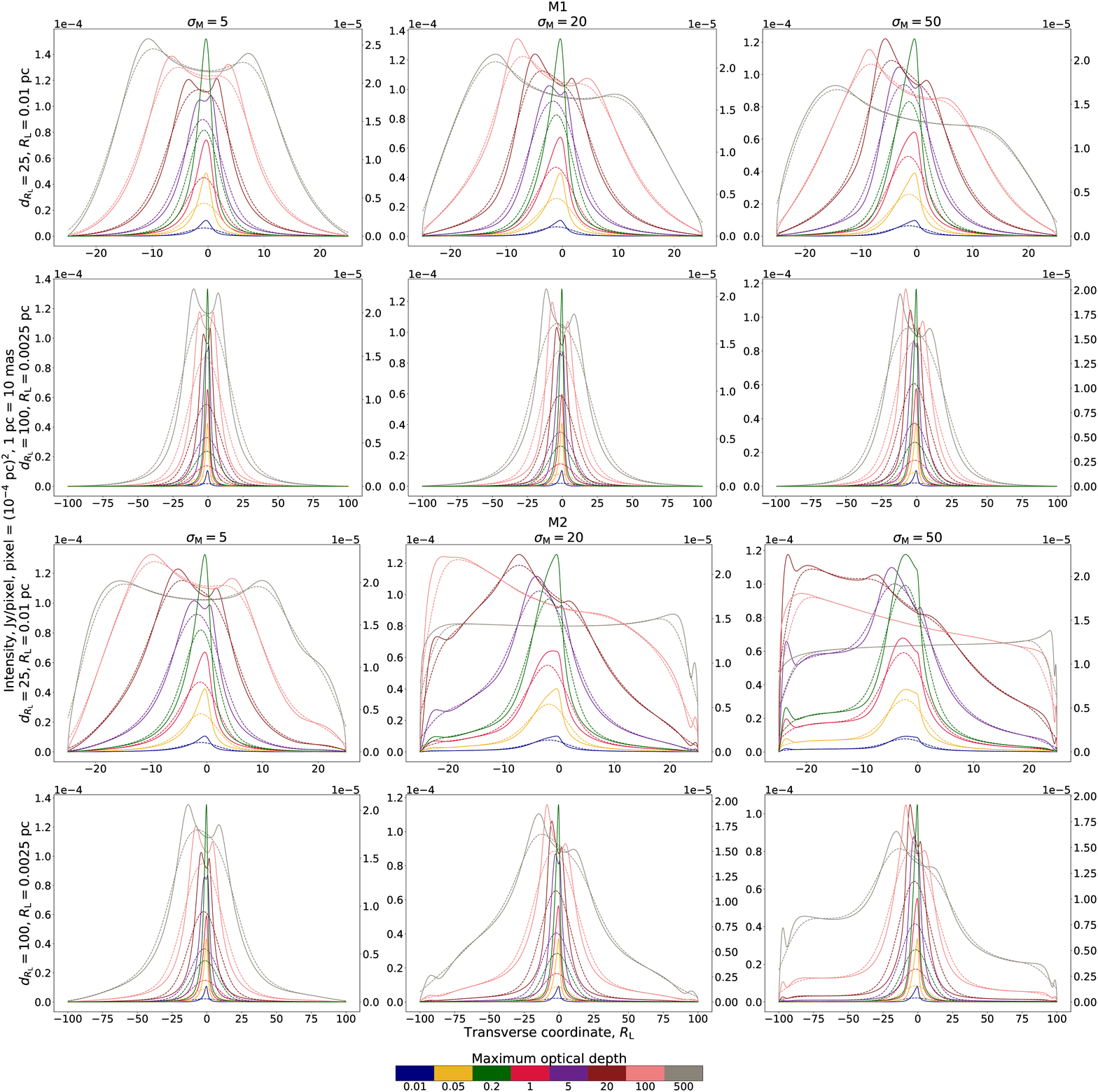}
    \caption{The same as in Figure \ref{fig:blazars_spin}, but at the viewing angle $15^{\circ}$. The physical parameters for this figure are summarized in Tables \ref{tab:radio_spin} and \ref{tab_app:radio_spin}.}
    \label{fig:radio_spin}
\end{figure*}

In Figures \ref{fig:blazars_spin} and \ref{fig:radio_spin} we present the intensity profiles plot sets in the `$R_{\rm L}$ comparison' mode: the geometrical jet width is set to the same value $d_\mathrm{jet}=0.5$~pc for every plot, so different $d_{R_{\rm L}}$ values relate to different $R_{\rm L}$ values. 
For the magnetic field lines threading the BH ergosphere, the condition of a maximum jet power $\Omega_\mathrm{F}=\Omega_\mathrm{H}/2$ can be used \citep{BZ-77}, although there are some indications of $\Omega_\mathrm{F}$ being much lower than the BH angular velocity $\Omega_\mathrm{H}$ \citep{Nokhrina19, Kino22}. In this case,
the light cylinder radius can be related to the dimensionless BH spin $a*$ according to the expression \citep{Mertens16, NKP20_r2}
\begin{equation}
    |a*| = \frac{8 a}{1 + 16a^2},\ a = \frac{r_{\rm g}}{R_{\rm L}},
\end{equation}
where $r_{\rm g}$ is the gravitational radius. We plot the dependence of $a*$ on $R_{\rm L}$ in Figure \ref{BHspin_vs_RL} having chosen M87 as a reference example.
We regard only the absolute values of BH spins as the positive and negative values and indistinguishable within our approach.
Down to $4 r_{\rm g}$ the smaller light cylinder radii correspond to higher BH spins, so we compare the jets from BHs with different spins while comparing $R_{\rm L}$ values. Thus, `$R_{\rm L}$ comparison' mode attitude aims to determine how wide or narrow transversely should be the central core to describe a particular given cross-section of the jet.

The plot sets are splitted in two major parts, upper and lower, dedicated to each of M1 and M2; $\sigma_{\rm M}$, i.e. terminal Lorentz factor, grows along the horizontal axis, and $d_{R_{\rm L}}$, i.e. BH spin, grows down the vertical axis. The plot sets are done for the viewing angles $5^{\circ}$ (Figure \ref{fig:blazars_spin}) and $15^{\circ}$ (Figure \ref{fig:radio_spin}) to capture common features of intensity profiles for quasars and radio galaxies. We show both the ideal, immediate result of the radiative transfer and the smoothed plots, convolved with the round beam of $0.5$ mas chosen so to be the order of magnitude less than the preset jet width. The intensity values are expressed in the unities of Jy/pixel, where the pixel is a square with sides of $10^{-4}$ pc. This pixel is chosen as it is intrinsic in our calculations. In this case, the beam  the order of magnitude less than the preset jet width would contain $\propto 10^6$ pixels.

There is a strong difference in the intensity amplitude for the plots with different optical depth $\tau_{\rm max}$. So, we plot them at the different scales, with the values for $\tau_{\rm max} \geqslant 1$ indicated on the left and for $\tau_{\rm max} < 1$ on the right axis.

\textit{1. BL Lacs/quasars/blazars and radio galaxies systematics.} 

Comparing Figures \ref{fig:blazars_spin} and \ref{fig:radio_spin} we observe, that, on average, the quasar intensity distribution is wider than the one in radio galaxies. This is a consequence of boosting of emission from a smaller part of a jet for a larger observational angle.

\textit{2. Plasma velocity effects.} 

The bulk Lorentz factor of a flow in a particular jet model impacts the intensity profile symmetry. Lorenz factors in our plots depend on the initial magnetization $\sigma_\mathrm{M}$, with slower jets presented in the left column and the faster jets in the right. Moving from lower to higher Lorentz factors, we observe, that the almost symmetrical profiles on the left become strongly asymmetrical on the right. This effect is again due to boosting the plasma emission by the flow with a toroidal velocity of the order of $0.1\;c$. We also checked the dependence of the intensity profile asymmetry on the viewing angle in the plasma frame that changes with the increase of the Lorentz factor (i.e. relativistic aberration) by considering profiles with different $\sigma_{\rm M}$ but artificially setting the zero toroidal velocity and found insignificant effect. The symmetry of an intensity profile depends on both the velocity and the observational angle (compare the second column, first line panel in Figures \ref{fig:blazars_spin} and \ref{fig:radio_spin}): the condition $\theta_\mathrm{obs}<1/\Gamma$ demarcates the asymmetric from symmetric profiles. 

High Lorentz factors in a jet (panels in the right columns) also lead to the overall offset of intensity towards the approaching part of a jet. In fact, in this case we should observe only one edge of a jet. As in one half of the jet the toroidal field projection is the same, in the polarization measurements this effect should be seen as a Faraday rotation measure (RM) with only one sign. Such RMs are observed in e.g. 3C 78, 4C +40.24, S5 0212+73, 3C 111, DA 406 \citep{Gabuzda2017}. Thus, observations of one sign RM do not necessarily imply the absence of the toroidal magnetic field component in a jet. They could be explained by the shift in emission intensity maximum for high bulk flow Lorentz factors --- of the order of $20$ and more, while the opposite side of a jet with other sign of magnetic field along a line of sight is not seen due to much lower intensity.

In principle, the influence of the high Lorenz factors on the transverse jet asymmetry could be tested with VLBI observations. However, there are several important issues to consider. First, as shown in \cite{017MNRAS.468.4992P} on a MOJAVE sample a true jet geometry in a considerable fraction of AGNs appears only after stacking single epoch maps over several years. Second, VLBI kinematic measurements could estimate not the true plasma, but the pattern speed, although there are other methods, which require simultaneous multi-frequency VLBI observations \citep[e.g.][]{kutkin19,core_shift_var}.

\textit{3. Optical depth effects.} 

\begin{figure}
    \centering
    \includegraphics[width = 0.96\columnwidth]{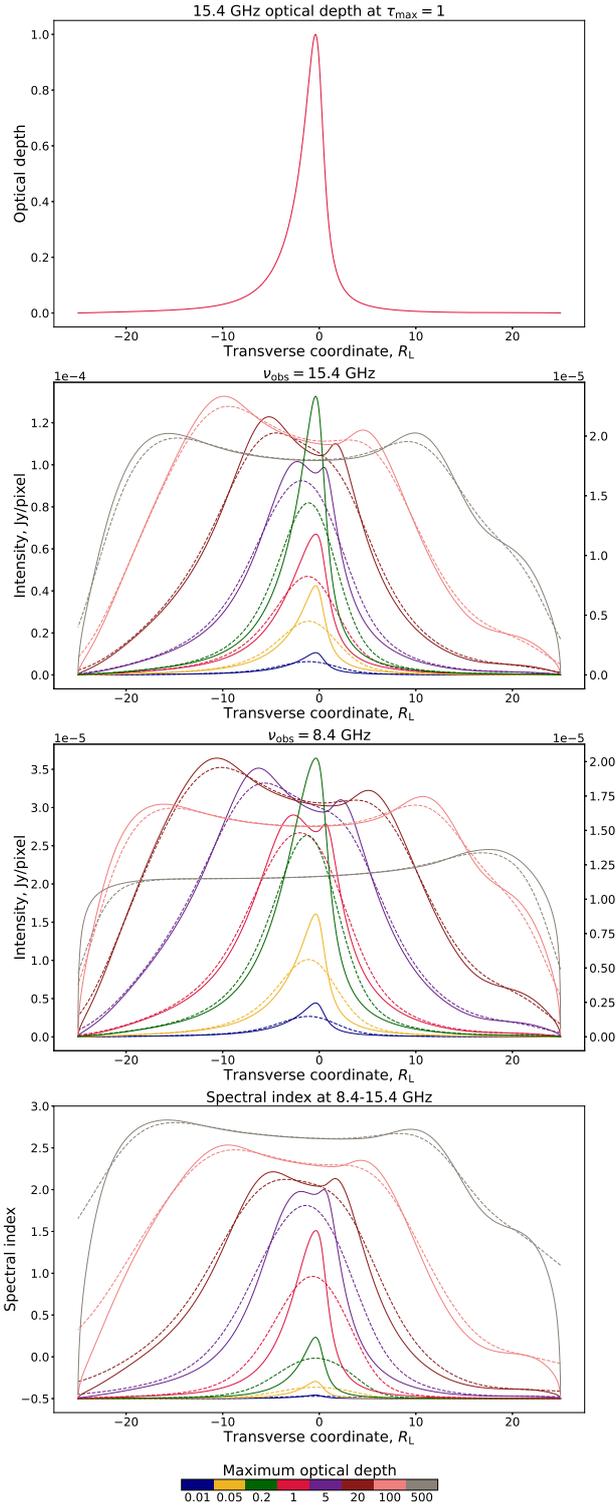}
    \caption{Within $({\rm M2}, \sigma_{\rm M} = 5, d_{R_{\rm L}} = 25)$ at $15^{\circ}$, the $15.4$ GHz optical depth profile at $\tau_{\rm max} = 1$ (so it corresponds to the $\tau_{\rm max} = 1$ $15.4$ GHz profiles of the same colour), the intensity profiles at $15.4$ (the same as in Figure \ref{fig:radio_spin}) and $8.4$ GHz and the corresponding spectral index. The solid lines are the ideal intensity profiles/the spectral index calculated based on the ideal intensity profiles, the dashed lines are the convolved intensity profiles/the spectral index calculated based on the convolved intensity profiles. The colours correspond to the maximum optical depth over the cross-section, $\tau_{\rm max}$. $\tau_{\rm max} \leqslant 1$ are plotted at the left y-axis, and $\tau_{\rm max} < 1$ are plotted at the right y-axis. The convolving beam is round with the FWHM equal to $0.1$ of the cross-section width ($0.5$ mas).}
    \label{fig:spectral_review}
\end{figure}

Maximum optical depth in Figures \ref{fig:blazars_spin} and \ref{fig:radio_spin} is indicated by the colour. Due to substantial stratification of the jet, the averaged over the cross-section optical depth is noticeably lower than its maximum. For example, for the $d_{R_{\rm L}} = 25$ cross-sections it is the order of magnitude lower. This means, that maroon curve should roughly correspond to the radio core region.  Though the maximum-to-averaged ratio varies from model to model, using maximum values uniformly for all models makes sense for well-resolved sources. In this case, strong transverse gradients in optical depth can manifest themselves, for example, in transverse spectral index gradients. The illustration for the possible spectral index gradients is given in Figure \ref{fig:spectral_review}. In the figure, we choose the model $({\rm M2}, \sigma_{\rm M} = 5, d_{R_{\rm L}} = 25)$ at $15^{\circ}$, and we plot i. the $15.4$ GHz optical depth profile at $\tau_{\rm max} = 1$ to give the feel of transverse optical depth behaviour, ii. the intensity profiles at $15.4$ and $8.4$ GHz and iii. the corresponding spectral indices. It is seen that e.g. the spectral index of approximately $1$ or $2$ in the central part of the jet can be accompanied by optically thin jet edges (see the plots for $\tau_{\rm max} = 1; 5; 20$).

For optically thin regimes, the emission from the central core dominates the total jet emission. With the growth of optical depth, while the difference between brightened central core and weakly radiating edges becomes less prominent, the net radiation increases drastically. In most cases sufficient optical thickness leads to forming the profile with two maximums (the maximum optical depth of $5$ or more), and their profoundness varies with the symmetry. In the observations, the larger optical depth corresponds to the lower observational frequency. Though the dependence on the observational frequency and total magnetic flux are not the same, qualitatively, they are similar. Thus, we predict that the change from a spine-brightened intensity profile to a limb-brightened profile with lowering the observational frequency occurs in low magnetized jets from the BHs with high spins. This could be the case of the observed effect in 3C 273 \citep{Bruni2021}.

On the other hand, in the highly asymmetric intensity profiles optical thickening can result in the change of the brightened edge from one to another (the middle and right panels in the third row corresponding to high magnetization and low BH spin in both Figures \ref{fig:blazars_spin} and \ref{fig:radio_spin}). It has the same nature as spine-to-limb brightening transition: the stronger emitting region also absorbs stronger, so at sufficient optical depth it becomes less bright than the region that emits weaker and absorbs weaker. However, this effect is observed only in M2 model due the presence of a slower sheath around a fast spine. The observational constraints of such effect should help make further conclusions about the presence of a slow sheath and a relatively large light cylinder radius.

\textit{4. Impact of a slow sheath presence.} 

In M2 model the Lorentz factor goes to unity at the boundary, so the outer region of the jet is brightened in highly magnetized jets (the middle and right panels in the third line in Figure \ref{fig:radio_spin}) as the radiation of outer part of a fast spine is suppressed due to boosting. This effect is hinted in $5^{\circ}$ profiles, but more pronounced at the viewing angle $15^{\circ}$. We interpret it as the possible explanation for forming the triple-peaked intensity structure. We expect this structure to be pronounced in highly magnetized jets from the BHs with large enough light cylinder radius and observed at considerably large viewing angles. Particularly, M87 jet is observed at the viewing angle of $17.2 \pm 3.3$ according to \citet{Mertens16}, and it is likely to possess the large light cylinder radius \citep{Nokhrina19}. The said conditions are required for the symmetry and for the possibility to produce a clear effect in the optically thin jets. 

We also observe (Figure \ref{fig:radio_spin}, panels in the right column), that the emission de-boosting due to lower velocity towards the jet boundary alone does not lead to the limb brightening, but only to the intensity flattening (see, for example, Figure \ref{fig:radio_spin}, the purple, green, red curves in the panel in the right column, third line). The flat parts of these curves correspond to already lower Lorentz factor of a bulk plasma motion. The lower plasma number density and smaller magnetic field amplitude in this part of a jet leads to fall in emissivity. Combination of smaller emissivity and emission de-boosting leads to a flat intensity profile. However, as the plasma number density grows closer to a jet boundary \citep{CBP19}, together with a low velocity it results in a bump in intensity.   

Thus, we conclude, that one can expect the triple-peaked structure across a jet if i. there is a central core manifesting itself the best at the moderate optical depths; ii. the optically thin jet edges with slow velocity and large particle number density. The latter can be result of matter entrainment from the ambient medium, for example, disk wind \citep{Araudo11}. 

The intensity profiles peaks are somewhat smoothed due the convolution. They are drawn in dashed lines in the same figures. For the chosen beam, the low-scale stratified patterns disappear.
\subsection{`Evolution' mode}
\label{ss:evolution_mode}

\begin{figure*}
    \centering
    \includegraphics[width = \textwidth]{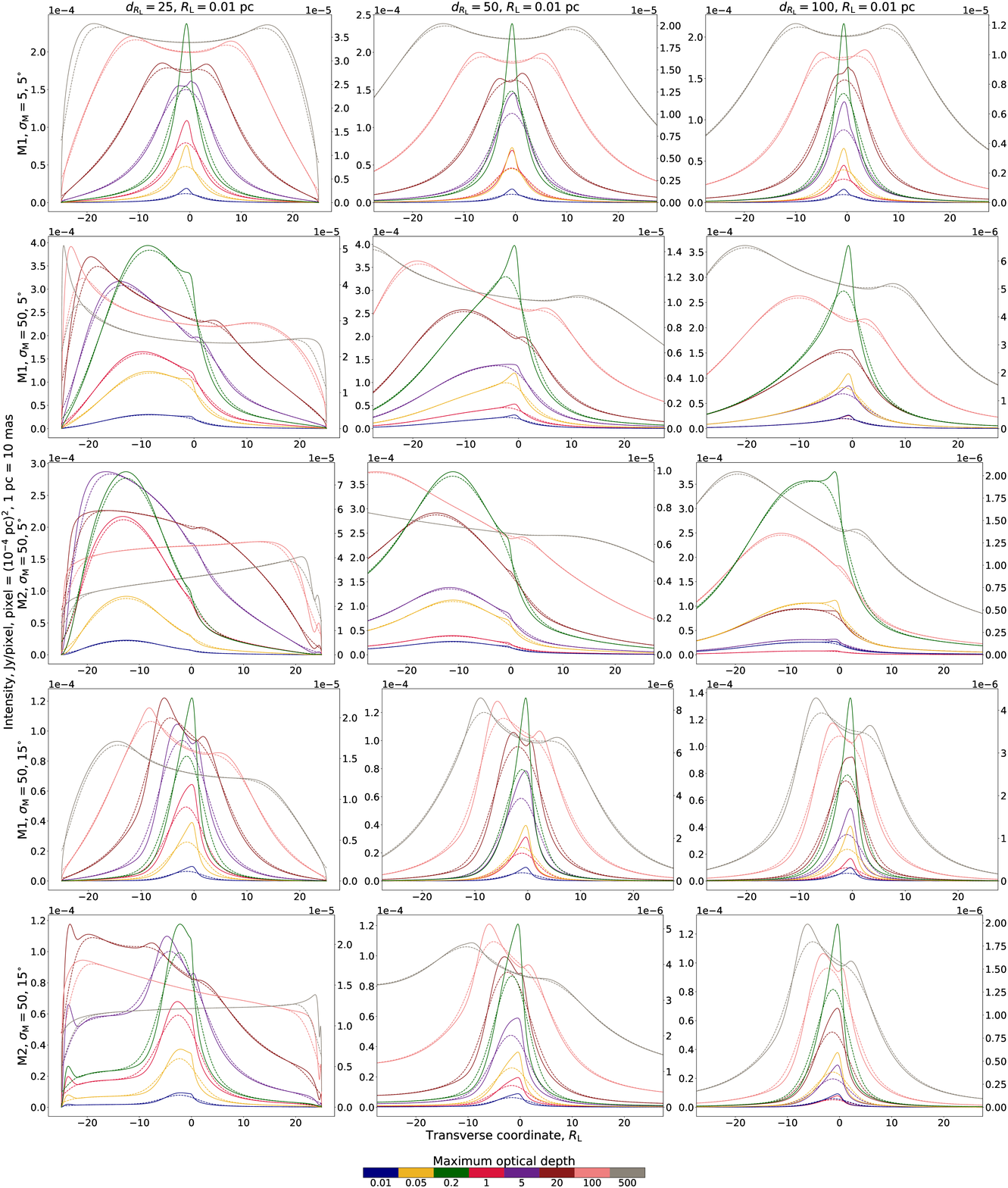}
    \caption{The ideal (solid lines) and convolved (dashed lines) intensity profiles describing the jet at different distances from the central engine; the distance grows from the left to the right; each row corresponds to the particular combination of the model (M1 or M2), initial magnetization $\sigma_{\rm M}$ and viewing angle. The colours correspond to the maximum optical depth over the cross-sections in the first column, $\tau_{\rm max}$. $\tau_{\rm max} \leqslant 1$ are plotted at the left y-axis, and $\tau_{\rm max} < 1$ are plotted at the right y-axis. The convolving beam is round with the FWHM equal to $0.1$ of the closest (the first column) cross-section width ($0.5$ mas). The physical parameters for this figure are summarized in Table \ref{tab_app:evolution}.}
    \label{fig:evolution}
\end{figure*}

In the previous subsection, we explored the typical intensity behaviour across a jet for different physical scenarios. In this subsection we want to explore how for every given model the intensity profile closer to the jet base changes as we move along a jet. 

The profiles in Figure \ref{fig:evolution} show how the given $d_{R_{\rm L}} = 25$ plot is supposed to evolve up to the widths of $d_{R_{\rm L}} = 50$ and $100$. The distance from the central engine grows from left to right, each row corresponds to the particular combination of the model (M1 or M2), initial magnetization $\sigma_{\rm M}$ and viewing angle. In this figure, the coloured panel showing the maximum optical depth corresponds to the plot colours at the panels in the first column only. As we move from the panel in the first column to the corresponding panel in the second and then third columns, we keep the model parameters, the observational frequency, and we designate it by using the same colour. For example, three grey curves in the upper row correspond to the same jet model, but for three cross-cuts at different distances from the jet base. As there are no edge effects at larger distances (at $d_{R_{\rm L}} = 50, 100$), the horizontal axis is cut at $25$ $R_{\rm L}$ to make clearer what happens on the same transverse scales. 

For the models with lower initial magnetization and relatively small bulk Lorentz factor (the first row in Figure \ref{fig:evolution}), we again observe symmetric intensity profiles and a transition from double-peaked to single-peaked structure as for the maroon line in two left panels for M1.

We propose that the jet stratification may influence the apparent change in a jet position angle (PA) along a jet. We see in the second and third rows of Figure \ref{fig:evolution}, corresponding to $\sigma_{\rm M} = 50$ and viewing angle $5^{\circ}$ in M1 and M2, green (optically thin) curve, that the intensity peak shifts along a jet noticeably. So, even for a straight jet we would observe a gradual shift in a jet PA due to effects of a jet stratification and optical depth. The same effect, though weaker, occurs for larger viewing angle (the fourth and fifth rows). For M87, Nikonov et al. submitted observe a slight change of the jet PA.

Additionally we should note, that, for a optically thin jet the intensity amplitude mostly declines as we move to a larger distances. While for a optically thick jet the intensity amplitude mostly maintains the same order of magnitude. 

\subsection{Intensity amplitudes and spectral flux}
\label{ss:total_flux}

The characteristic total spectral flux in almost all the presented above figures is much larger, than typical values, reported by the MOJAVE catalogue\footnote{\url{https://www.cv.nrao.edu/MOJAVE/index.html}} \citep{MOJAVE_XV}. Typical intrinsic brightness temperature is reported to be $(4.1 \pm 0.6) \times 10^{10}$ K \citep{MOJAVE_XIX}, which is recalculated for $15.4$ GHz and e.g. the round beam of $1$ mas as approximately $8$ Jy/beam. To compare, as we used the MOJAVE frequency $15.4$ GHz to construct the plots, consider the maximums of intensity in our $\tau_{\rm max} = 1$ plots. The maximum intensity amplitude on our plots is of the order of a few $10^{-4}$~Jy/pixel. Having a beam size of $10^6$ pixels, we obtain for a spectral flux values of the order of $100$~Jy --- several orders of magnitude higher, than the observed values in a jet core region. Only the most optically thin regimes have a spectral flux of a few Jy. 

A jet with a structure, obtained self-consistently within ideal axisymmetric stationary MHD approach, has a central core --- a dense particle-dominated part with the strongest magnetic field. This feature is robust. However, we observe, that in all the regimes, except for the most optically thick, it is the radiation from this core that dominates the overall jet emission. 
Thus, all our models have an excess flux as a consequence of accounting for an emission of a central core. 

We must conclude, that only a small fraction --- of an order of a percent --- of total particle number density are the emitting particles. The major part of the plasma is cold and does not emit. This is in agreement with the models of particle acceleration on shocks or in reconnection events \citep{SironiSpitkovsky2011, SironiSpitkovskyArons2013, SironiSpitkovsky2014}. In this case all the major effects of a jet stratified structure, described above in \autoref{ss:BHspin_mode}, are present, but with the total spectral flux values, closer to the observed ones.

\begin{table*}
	\centering
	\begin{tabular}{c|c|c|c|c|c|c|c|c|c}
\hline 
		 M & $\sigma_{\rm M}$ & $d_{R_{\rm L}}$ & \begin{tabular}[x]{@{}c@{}}$\Psi_0$, \\ $10^{34}$ G cm$^2$\end{tabular} & \begin{tabular}[x]{@{}c@{}}$W_{\rm jet}$, \\ $10^{44}$ erg s$^{-1}$ \end{tabular} & \begin{tabular}[x]{@{}c@{}} $n_{\rm max}^{\prime}$, \\ cm$^{-3}$\end{tabular} & \begin{tabular}[x]{@{}c@{}} $n_{\rm average}^{\prime}$, \\ cm$^{-3}$\end{tabular} & \begin{tabular}[x]{@{}c@{}} $B_{\rm max}^{\prime}$, \\ G\end{tabular}  & \begin{tabular}[x]{@{}c@{}} $B_{\rm average}^{\prime}$,\\ G\end{tabular}  &  \begin{tabular}[x]{@{}c@{}}$I_{\rm max}$, \\ $\mu$Jy/pixel \end{tabular}\\ \hline 
		M1 & 5 & 25 & 1.41 & 0.80 & 1453 & 156 & 0.055 & 0.014 & 79.35\\ 
		M1 & 20 & 25 & 2.51 & 2.52 & 807 & 91 & 0.069 & 0.020 & 103.92\\ 
		M1 & 50 & 25 & 4.26 & 7.25 & 599 & 71 & 0.075 & 0.025 & 161.21\\ 
		M1 & 5 & 100 & 0.18 & 0.20 & 3717 & 102 & 0.069 & 0.006 & 29.44\\ 
		M1 & 20 & 100 & 0.35 & 0.77 & 2237 & 64 & 0.086 & 0.009 & 40.85\\ 
		M1 & 50 & 100 & 0.65 & 2.68 & 1806 & 54 & 0.093 & 0.011 & 54.46\\ 
		M2 & 5 & 25 & 0.84 & 0.28 & 828 & 110 & 0.071 & 0.020 & 81.21\\ 
		M2 & 20 & 25 & 2.44 & 2.38 & 507 & 71 & 0.082 & 0.029 & 166.05\\ 
		M2 & 50 & 25 & 3.65 & 5.32 & 276 & 40 & 0.078 & 0.031 & 211.80\\ 
		M2 & 5 & 100 & 0.11 & 0.08 & 2195 & 69 & 0.088 & 0.009 & 35.05\\ 
		M2 & 20 & 100 & 0.38 & 0.93 & 1530 & 50 & 0.099 & 0.012 & 69.76\\ 
		M2 & 50 & 100 & 0.80 & 4.09 & 1288 & 44 & 0.103 & 0.015 & 99.94\\ 
 \hline 
	\end{tabular}
\caption{The physical parameters for the maximum optical depth of $\tau_{\rm max} = 1$ in Figure \ref{fig:blazars_spin}.}
 \label{tab:blazars_spin}
\end{table*}

\begin{table*}
	\centering
	\begin{tabular}{c|c|c|c|c|c|c|c|c|c}
\hline 
		 M & $\sigma_{\rm M}$ & $d_{R_{\rm L}}$ &  \begin{tabular}[x]{@{}c@{}}$\Psi_0$, \\ $10^{34}$ G cm$^2$\end{tabular} & \begin{tabular}[x]{@{}c@{}}$W_{\rm jet}$, \\ $10^{44}$ erg s$^{-1}$ \end{tabular} & \begin{tabular}[x]{@{}c@{}} $n_{\rm max}^{\prime}$, \\ cm$^{-3}$\end{tabular} & \begin{tabular}[x]{@{}c@{}} $n_{\rm average}^{\prime}$, \\ cm$^{-3}$\end{tabular} & \begin{tabular}[x]{@{}c@{}} $B_{\rm max}^{\prime}$, \\ G\end{tabular}  & \begin{tabular}[x]{@{}c@{}} $B_{\rm average}^{\prime}$,\\ G\end{tabular}  &  \begin{tabular}[x]{@{}c@{}}$I_{\rm max}$, \\ $\mu$Jy/pixel \end{tabular}\\ \hline
		M1 & 5 & 25 & 1.77 & 1.25 & 2276 & 244 & 0.069 & 0.017 & 45.17\\ 
		M1 & 20 & 25 & 3.21 & 4.12 & 1322 & 150 & 0.088 & 0.026 & 46.91\\ 
		M1 & 50 & 25 & 5.58 & 12.43 & 1028 & 122 & 0.098 & 0.033 & 49.38\\ 
		M1 & 5 & 100 & 0.22 & 0.32 & 5781 & 159 & 0.087 & 0.008 & 13.80\\ 
		M1 & 20 & 100 & 0.49 & 1.54 & 4485 & 128 & 0.122 & 0.012 & 14.55\\ 
		M1 & 50 & 100 & 0.83 & 4.36 & 2934 & 87 & 0.118 & 0.014 & 15.44\\ 
		M2 & 5 & 25 & 1.06 & 0.44 & 1298 & 172 & 0.089 & 0.025 & 46.90\\ 
		M2 & 20 & 25 & 3.17 & 4.02 & 855 & 120 & 0.106 & 0.038 & 55.05\\ 
		M2 & 50 & 25 & 5.75 & 13.18 & 685 & 100 & 0.122 & 0.050 & 59.14\\ 
		M2 & 5 & 100 & 0.14 & 0.13 & 3444 & 108 & 0.111 & 0.011 & 14.86\\ 
		M2 & 20 & 100 & 0.49 & 1.54 & 2523 & 82 & 0.127 & 0.016 & 16.70\\ 
		M2 & 50 & 100 & 1.05 & 7.00 & 2202 & 75 & 0.135 & 0.020 & 17.49\\ 
 \hline 
  average & & & & & 2400 & 130 & 0.106 & & \\
 BK & & & & & 11 & & 0.106 & & \\ \hline
	\end{tabular}
 \caption{The physical parameters for the maximum optical depth of $\tau_{\rm max} = 1$ in Figure \ref{fig:radio_spin}. The last but one line presents the averaged over the models magnetic field and particle number density, and the last line presents the Blandford-K\"onigl model prediction of particle number density for such magnetic field (see \autoref{s:discussion}).}
  \label{tab:radio_spin}
\end{table*}

\section{Other distributions of nonthermal electrons across a jet}
\label{s:heating}

Using the robust model for the central part of a jet, we obtained a dominance in jet emission of a central core or nearby jet parts. Double-peaked intensity profiles require extremely high optical depths or a combination of high magnetization and considerably large viewing angle. Triple-peaked intensity profiles require even stronger combination: high magnetization, high BH spin and larger viewing angle, as the symmetry is needed to highlight both of the jet edges. Additionally, the said peaks are not pronounced in comparison with the central core-defined peak. The intensity values are also notably higher than we expect for nearby sources, e.g. M87. So, it is likely that the jet transverse structure itself cannot describe the observations.
We conclude that the limb brightening can be reconstructed in our modelling only if a jet has a slow sheath with the same order of particle number density and magnetic field, as in the central core. Larger particle number density with small velocity can be connected with a disc wind \citep{BlP-82, Stone99, GL17, Lu23}. However, having the same amplitudes of a magnetic field as in the dense core may be problematic. In order to reconcile our findings with the observations, we consider different distributions of the nonthermal (i.e. emitting) plasma. 

\subsection{Spatial distribution of nonthermal particles}
\label{ss:spatial}

Acceleration of emitting plasma particles on instabilities may be expected preferably in the vicinity of a jet boundary due to developing instabilities \citep{McKinney2006, Chatterjee2019, Hardee2011}. The Model~1 may be favourable for the shear acceleration at the jet boundary, where the bulk motion Lorentz factor is the greatest across a jet, and  plasma with $\Gamma\gg 1$ contacts the ambient medium with non-relativistic bulk velocity \citep{Ostrowski1990, Ostrowski1998}. In order to account for these effects, 
we choose the simple Heaviside-like step function with two parameters to describe nonthermal plasma spatial distribution:
\begin{equation}
\label{lowering_function}
    f(r, k_{\%}, k_r) = \begin{cases}
        k_{\%}, r < k_r d, \\
        1, k_r d \leqslant r \leqslant d,
        \end{cases}
\end{equation}
$0 < k_{\%, r} < 1$, so $k_{\%}$ stands for the fraction of the particle number density in the central core that is emitting, and $k_r d$ is the radial coordinate, starting from which we assume that all the plasma emits. 
This function choice implies that the source of nonthermal electrons is due to interaction with the surrounding matter, 
while the central region does not undergo such interaction. Discontinuity of the function does not affect the qualitative results but we additionally linearised the transition between two areas, and it can be easily smoothed in other ways. In calculations, we put the renewed $k_e$ function into the general synchrotron emission expressions for emission (\ref{j}) and absorption (\ref{ae}) coefficients.

\begin{figure*}
\centering
\includegraphics[width=\textwidth, angle=0]{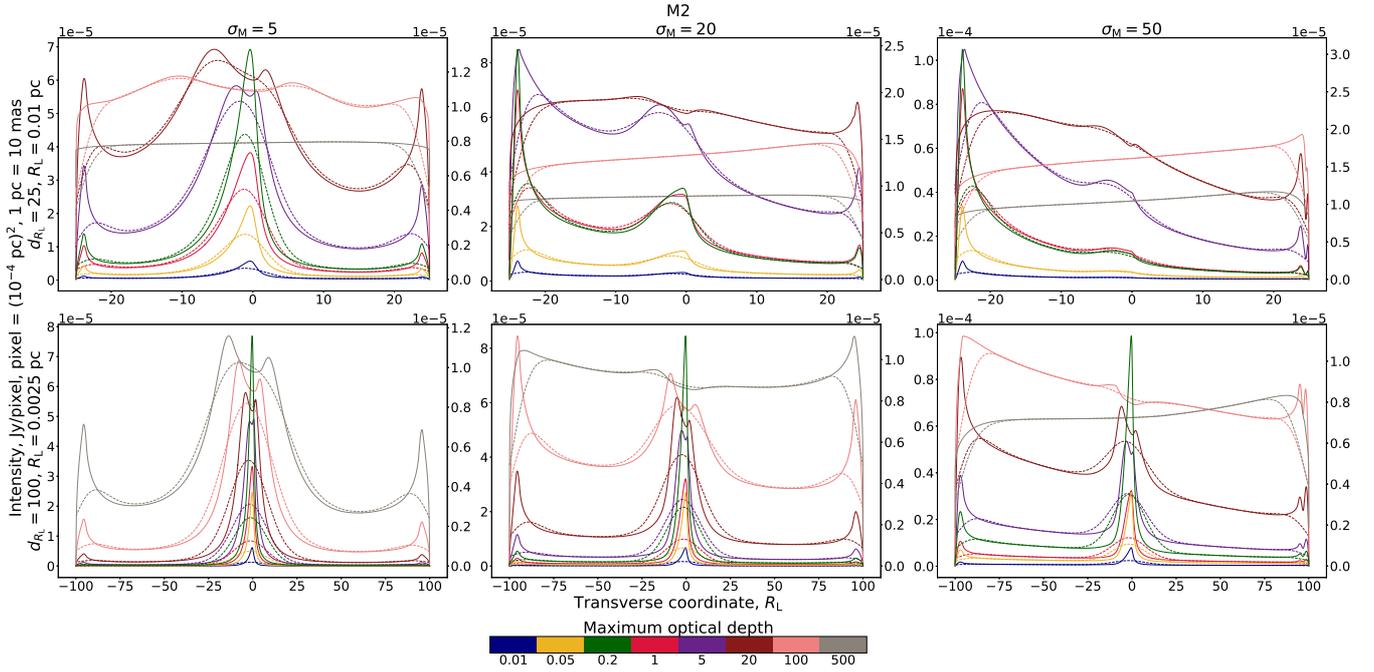}
\caption{The ideal (solid lines) and convolved (dashed lines) intensity profiles in M2 at the viewing angle $15^{\circ}$ for the nonthermal electron distribution defined in \eqref{lowering_function} with the parameters $k_{\%} = 0.01$, $k_r = 0.95$. The $R_{\rm L}$ values are chosen in the `$R_{\rm L}$ comparison' mode (the first and second row for each model describe the jet cross-section of the same geometrical width, $0.5$ pc, but with different light cylinder radii, see details in \autoref{ss:BHspin_mode}). The colours correspond to the maximum optical depth over the cross-section, $\tau_{\rm max}$. $\tau_{\rm max} \leqslant 1$ are plotted at the left y-axis, and $\tau_{\rm max} < 1$ are plotted at the right y-axis. The convolving beam is round with the FWHM equal to $0.1$ of the cross-section width ($0.5$ mas). The physical parameters for this figure are summarized in Tables \ref{tab:radio_BHspin_l0.01l0.95} and \ref{tab_app:radio_BHspin_l0.01l0.95}.}
\label{fig:radio_BHspin_l0.01l0.95}
\end{figure*}

\begin{table*}
	\centering
	\begin{tabular}{c|c|c|c|c|c|c|c|c|c}
 \hline
		 M & $\sigma_{\rm M}$ & $d_{R_{\rm L}}$ & \begin{tabular}[x]{@{}c@{}}$\Psi_0$, \\ $10^{34}$ G cm$^2$\end{tabular} & \begin{tabular}[x]{@{}c@{}}$W_{\rm jet}$, \\ $10^{44}$ erg s$^{-1}$ \end{tabular} & \begin{tabular}[x]{@{}c@{}} $n_{\rm max}^{\prime}$, \\ cm$^{-3}$\end{tabular} & \begin{tabular}[x]{@{}c@{}} $n_{\rm average}^{\prime}$, \\ cm$^{-3}$\end{tabular} & \begin{tabular}[x]{@{}c@{}} $B_{\rm max}^{\prime}$, \\ G\end{tabular}  & \begin{tabular}[x]{@{}c@{}} $B_{\rm average}^{\prime}$,\\ G\end{tabular}  &  \begin{tabular}[x]{@{}c@{}}$I_{\rm max}$, \\ $\mu$Jy/pixel \end{tabular}\\ \hline 
		M2 & 5 & 25 & 3.32 & 4.40 & 1507 & 67 & 0.281 & 0.079 & 27.26\\ 
		M2 & 20 & 25 & 8.49 & 28.75 & 1698 & 46 & 0.284 & 0.101 & 34.18\\ 
		M2 & 50 & 25  & 10.06 & 40.39 & 919 & 19 & 0.214 & 0.087 & 41.23\\ 
		M2 & 5 & 100  & 0.45 & 1.27 & 436 & 22 & 0.350 & 0.035 & 8.41\\ 
		M2 & 20 & 100  & 1.55 & 15.35 & 915 & 21 & 0.400 & 0.050 & 9.89\\ 
		M2 & 50 & 100  & 3.30 & 69.41 & 1581 & 22 & 0.424 & 0.063 & 12.21\\ 
 \hline 
  average & & & & & 1180 & 33 & 0.326 & & \\
 BK & & & & & 106 & & 0.326 & & \\ \hline
	\end{tabular}
 \caption{The physical parameters for the maximum optical depth of $\tau_{\rm max} = 1$ in Figure \ref{fig:radio_BHspin_l0.01l0.95}.  The last but one line presents the averaged over the models magnetic field and particle number density, and the last line presents the Blandford-K\"onigl model prediction of particle number density for such magnetic field (see \autoref{s:discussion}).}
 \label{tab:radio_BHspin_l0.01l0.95}
\end{table*}

The resulting intensity profiles with nonthermal particle spatial distribution $n(r)f(r,\,k_{\%},\,k_r)$ are presented in
Figure \ref{fig:radio_BHspin_l0.01l0.95}. We chose M2 and $k_{\%} = 0.01$, $k_r = 0.95$ for demonstrative purposes. At $1$ per cent of the emitting plasma in a jet dense core, the jets with higher $d_{R_{\rm L}}$ (smaller light cylinder radius; the second row) still possess the prominent central core, while for low $d_{R_{\rm L}}$ (the first row) such suppression of the central core emission is enough to obtain the comparable intensity amplitudes of the center and the edges. We obtain more symmetric distributions for low magnetization and/or smaller light cylinder radius. Thus, we predict that at least one of this conditions is fulfilled for the jets with symmetric triple-peaked intensity profiles: either relatively low Lorentz factors of a flow, or small light cylinder radius (which may correspond to high BH spin). 
 
Therefore, we predict that the interaction with the ambient medium is a key process for the limb brightening appearance in an optically thin jets. In the same time, the central core emission suppression moderates the intensity values in average (compare Tables \ref{tab:radio_spin} and \ref{tab:radio_BHspin_l0.01l0.95}).

Thus, excluding the spine emission is a promising mechanism for producing the brightened areas close to the jet boundaries, though the particular realisation and extent of the effect are under question. We note that other similar mechanisms of excluding the emission from the central region are considered, e.g. in \citet{CruzOsorio2022} and \citet{Fromm2022} the emitting jet sheath and non-emitting jet spine are separated basing on the local magnetization.

\subsection{Ohmic heating}
\label{ss:Ohm}

In this subsection we want to show how other assumptions on the nonthermal plasma spatial distribution affects the intensity profiles of synchrotron radiation from the stratified jets.

It was shown by \citet{Lyutikov2005}, that the observed jet linear polarization for specially chosen magnetic field and particle number density distributions across a jet is consistent with the following distribution: $k_e \sim j'^2$. Here $j'$ is an electric current density in the plasma proper frame, which essentially implies Ohmic heating. Since
\begin{equation}
    j_z = \frac{c}{4\pi r} \frac{\partial (r B_{\varphi})}{\partial r},
\end{equation}
\begin{equation}
    j_{\varphi} = -\frac{c}{4\pi} \frac{\partial B_z}{\partial r},
\end{equation}
the electric current peaks in the central core as the magnetic field decreases abruptly with $r$. Within M2, it peaks as well at the jet boundary due to the poloidal magnetic field rapid vanishing \citep{CBP19}.

\begin{figure*}
\centering
\includegraphics[width=\textwidth, angle=0]{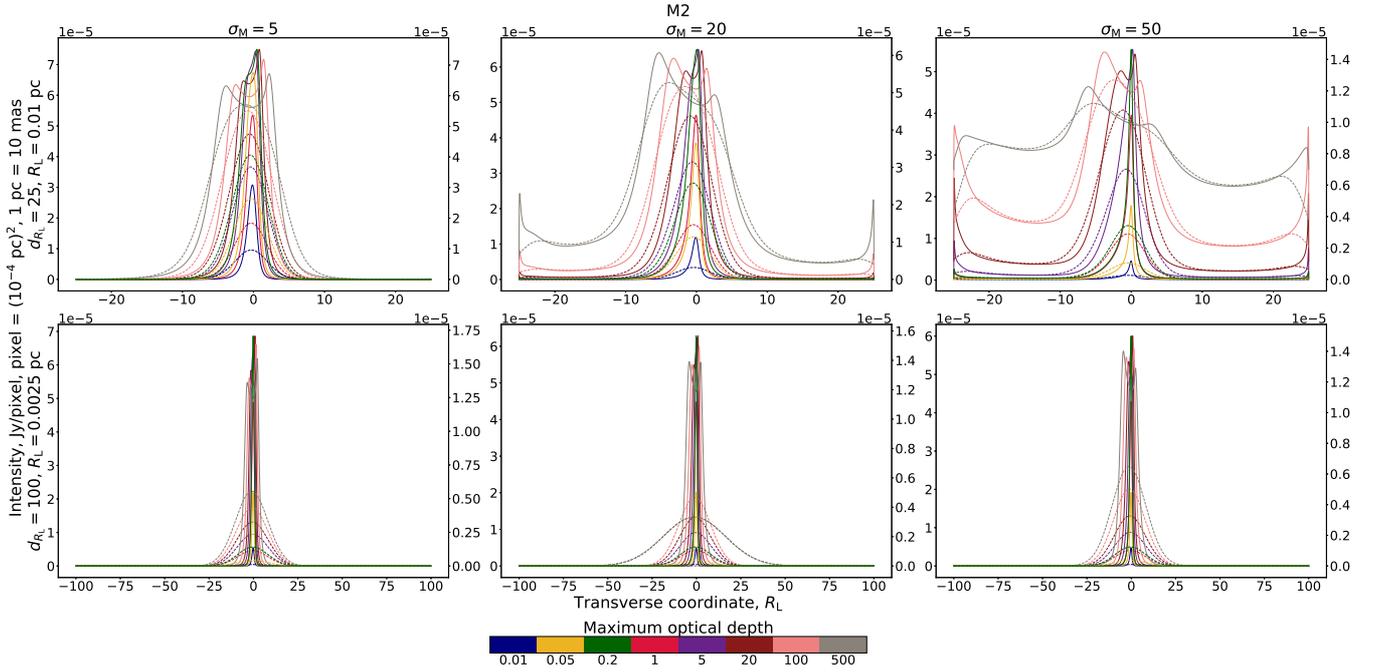}
\caption{The ideal (solid lines) and convolved (dashed lines) intensity profiles in M2 at the viewing angle $15^{\circ}$ for Ohmic heating ($k_e \sim j'^2$). The $R_{\rm L}$ values are chosen in the `$R_{\rm L}$ comparison' mode (the first and second row for each model describe the jet cross-section of the same geometrical width, $0.5$ pc, but with a different light cylinder radii, see details in \autoref{ss:BHspin_mode}). The colours correspond to the maximum optical depth over the cross-section, $\tau_{\rm max}$. $\tau_{\rm max} \leqslant 1$ are plotted at the left y-axis, and $\tau_{\rm max} < 1$ are plotted at the right y-axis. The convolving beam is round with the FWHM equal to $0.1$ of the cross-section width ($0.5$ mas). The physical parameters for this figure are summarized in Tables \ref{tab:radio_BHspin_Ohmic} and \ref{tab_app:radio_BHspin_Ohmic}.}
\label{fig:radio_BHspin_Ohmic}
\end{figure*}

To ensure the emitting particle number density does not exceed the total number density obtained within the MHD modelling, we set
\begin{equation}
    k_e = n_{p, \gamma_{\rm min}}(r = 0) (j'/{\rm max}(j'))^2.
    \label{Ohmic}
\end{equation} 
The intensity profiles assuming Ohmic heating are presented in Figure \ref{fig:radio_BHspin_Ohmic}.
We see that due to the very narrow $j'$ distribution, the central peaks are much more narrow than in case of the emission from the central dense core plasma in \autoref{s:intensity}. This
narrow intensity peak, being smoothed by the convolution, provides the sufficiently lower intensity amplitude than in the previous models (compare Tables \ref{tab:radio_spin} and \ref{tab:radio_BHspin_Ohmic}). The effect strengthens as the initial magnetization $\sigma_{\rm M}$ grows since in a highly magnetized jet the magnetic field derivatives are stronger across the jet, and the electric current density has even more narrow spatial distribution near the jet axis and at the very boundary. 

In the case of high magnetization and considerable optical depth ($\geqslant 20$) the rapid $j'$ growth at the jet boundary manifests itself in the intensity peak (the third column in the first row in Figure \ref{fig:radio_BHspin_Ohmic}). Thus, the Ohmic heating can possibly account for the observed triple-peaked intensity profiles. Another very important consequence of such nonthermal particle distribution is the clear transverse symmetry that manifests in the brightening of both edges even for high bulk flow Lorentz factors.

\begin{table*}
	\centering
	\begin{tabular}{c|c|c|c|c|c|c|c|c|c}
 \hline
		 M & $\sigma_{\rm M}$ & $d_{R_{\rm L}}$ &  \begin{tabular}[x]{@{}c@{}}$\Psi_0$, \\ $10^{34}$ G cm$^2$\end{tabular} & \begin{tabular}[x]{@{}c@{}}$W_{\rm jet}$, \\ $10^{44}$ erg s$^{-1}$ \end{tabular} & \begin{tabular}[x]{@{}c@{}} $n_{\rm max}^{\prime}$, \\ cm$^{-3}$\end{tabular} & \begin{tabular}[x]{@{}c@{}} $n_{\rm average}^{\prime}$, \\ cm$^{-3}$\end{tabular} & \begin{tabular}[x]{@{}c@{}} $B_{\rm max}^{\prime}$, \\ G\end{tabular}  & \begin{tabular}[x]{@{}c@{}} $B_{\rm average}^{\prime}$,\\ G\end{tabular}  &  \begin{tabular}[x]{@{}c@{}}$I_{\rm max}$, \\ $\mu$Jy/pixel \end{tabular}\\ \hline 
		M2 & 5 & 25 & 1.33 & 0.70 & 2048 & 65 & 0.112 & 0.032 & 18.33\\ 
		M2 & 20 & 25 & 4.49 & 8.03 & 1708 & 37 & 0.150 & 0.053 & 15.46\\ 
		M2 & 50 & 25 & 9.86 & 38.78 & 2016 & 21 & 0.209 & 0.085 & 11.07\\ 
		M2 & 5 & 100 & 0.17 & 0.19 & 5054 & 45 & 0.134 & 0.013 & 4.61\\ 
		M2 & 20 & 100 & 0.62 & 2.46 & 4029 & 31 & 0.160 & 0.020 & 4.27\\ 
		M2 & 50 & 100 & 1.36 & 11.81 & 3715 & 26 & 0.175 & 0.026 & 4.04\\ 
 \hline 
   average & & & & & 3100 & 38 & 0.157 & & \\
 BK & & & & & 25 & & 0.157 & & \\ \hline
	\end{tabular}
 \caption{The physical parameters for the maximum optical depth of $\tau_{\rm max} = 1$ in Figure \ref{fig:radio_BHspin_Ohmic}. The last but one line presents the averaged over the models magnetic field and particle number density, and the last line presents the Blandford-K\"onigl model prediction of particle number density for such magnetic field (see \autoref{s:discussion}).}
 \label{tab:radio_BHspin_Ohmic}
\end{table*}

\subsection{Emitting plasma equipartition}
\label{ss:equipartition}

Another possible emitting particle distribution discussed in \citet{Burbidge56, BlandfordKoenigl1979} is the equipartition between the emitting particles energy and magnetic field in the plasma proper frame $B'$, which leads to $k_e \sim B'^2$. We do not expect suppression of the central core emission in this case. Indeed, the poloidal magnetic field $B_{\mathrm{p}}$ is almost constant within the central core \citep{Beskin09}, while the toroidal component $B_{\varphi}$ reaches the same amplitude at the light cylinder radius. Beyond $R_\mathrm{L}$, both components fall with the conserved ratio $B_{\mathrm{p}}/B_{\varphi}=R_{\mathrm{L}}/r$ while plasma is relativistic \citep[e.g.,][]{Lyu09}. For the strongly magnetized accelerating flow $\Gamma\approx r/R_{\mathrm{L}}$, and both components are of the same order in the plasma proper frame \citep{Lyutikov2005}:
\begin{equation}
\frac{B'_{\mathrm{p}}}{B'_{\varphi}}=\Gamma\frac{B_{\mathrm{p}}}{B_{\varphi}}\approx 1.
\end{equation}
Further along the jet for $r>\sigma_{\mathrm{M}}R_{\mathrm{L}}$, $B'_{\varphi}$ dominates $B'_{\mathrm{p}}$. However, in both cases the amplitude $B'^2$ is maximal within the core, and no emission suppression is expected: the central core dominates the optically thin jet. Indeed, we carried out the calculations with
\begin{equation}
    k_e = \min \{B'^2 / (8\pi m c^2), n_{p, \gamma_{\rm min}}\}
\end{equation} 
and did not get any noticeable difference between the equipartition regime and all the jet plasma emitting.

\section{Discussion}
\label{s:discussion}
\begin{figure}
\centering
\includegraphics[width=\columnwidth, angle=0]{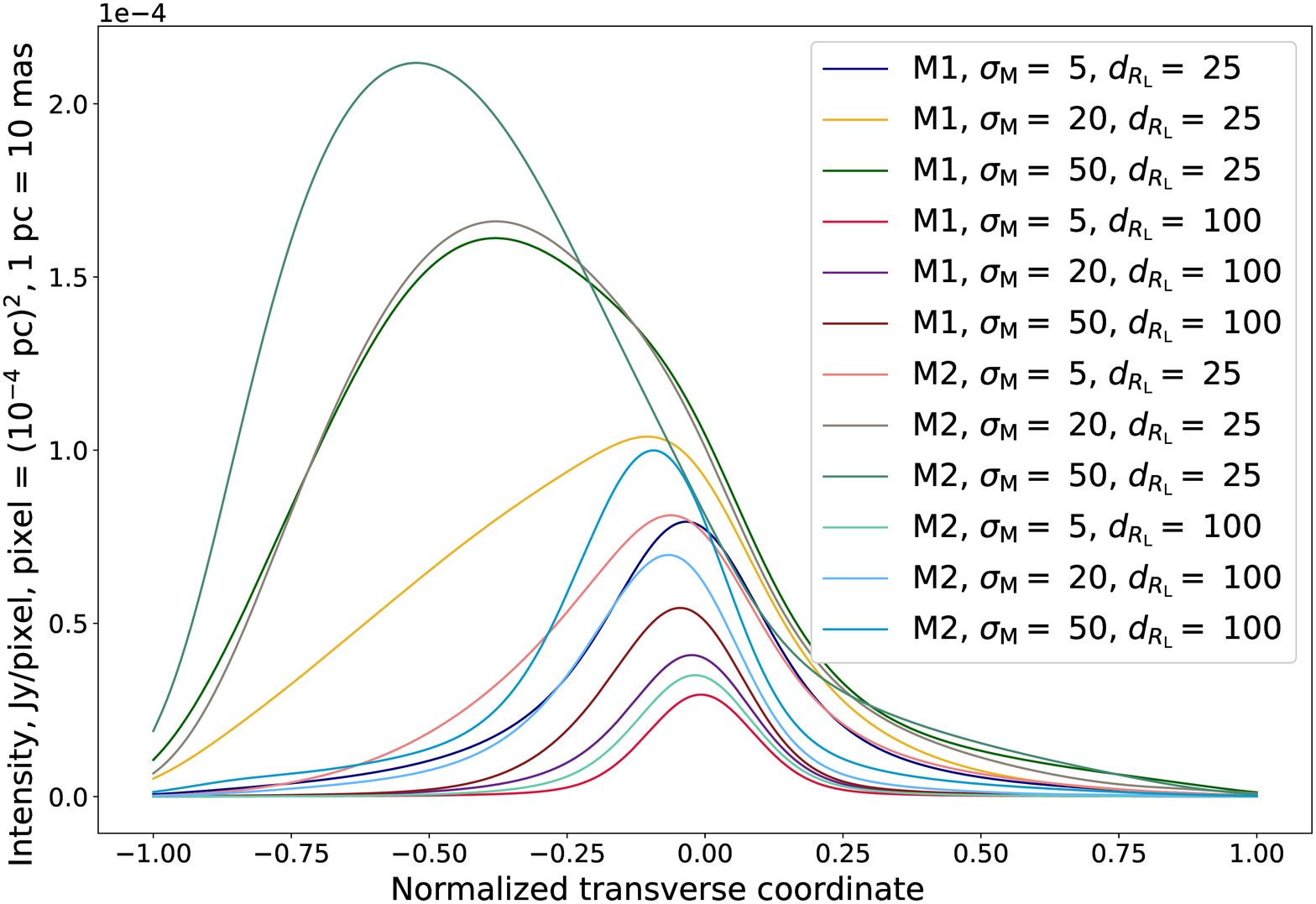}
\includegraphics[width=\columnwidth, angle=0]{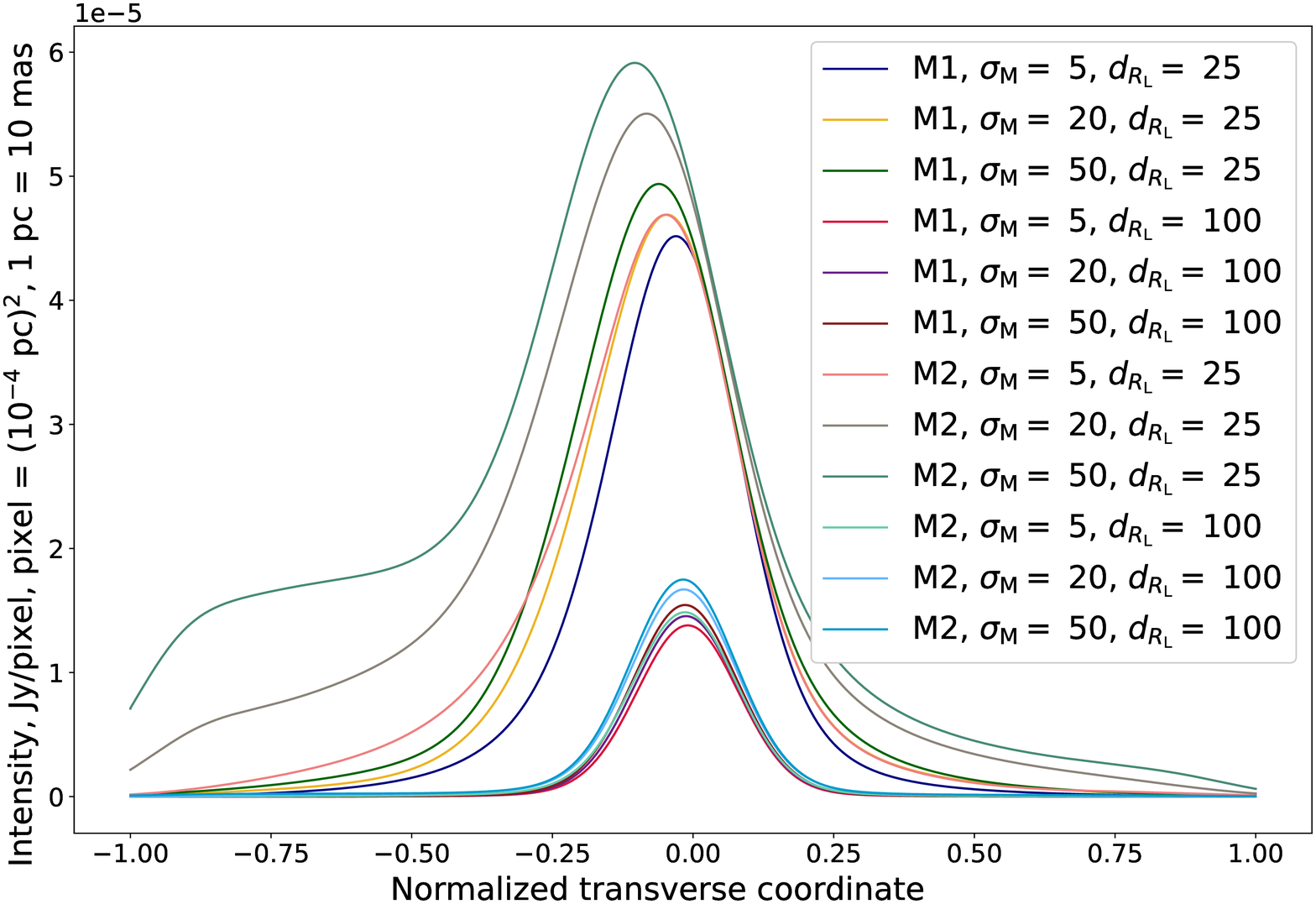}
\caption{The convolved intensity profiles at $\tau_{\rm max} = 1$ for the models with all jet plasma emitting according to a power-law distribution. The viewing angles are $5^{\circ}$ and $15^{\circ}$ for upper and lower plots correspondingly.}
\label{fig:blazars_radio_tau1}
\end{figure}

\begin{figure}
\centering
\includegraphics[width=\columnwidth, angle=0]{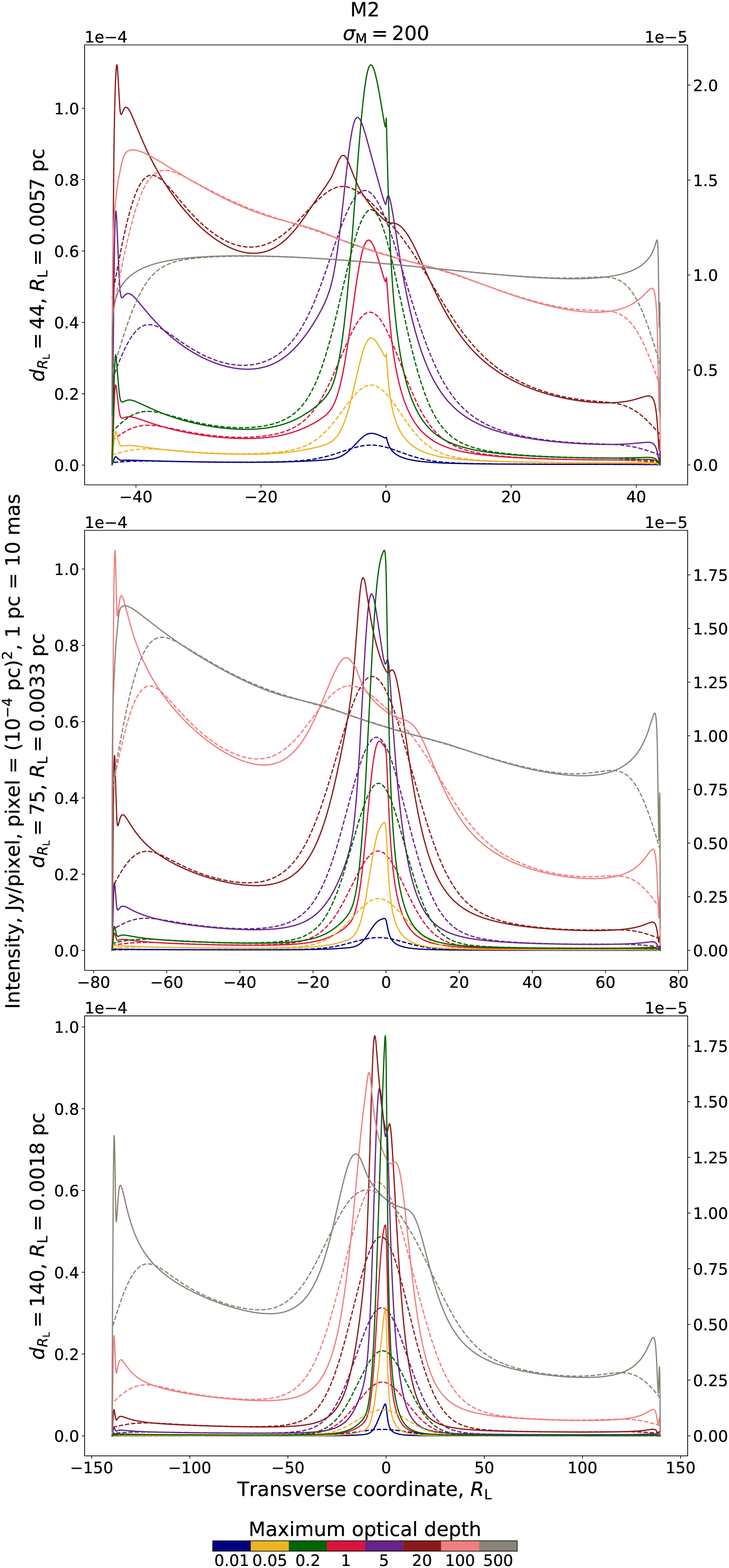}
\caption{The intensity profiles for the magnetization $\sigma_{\rm M} = 200$ in the M2 with all jet plasma emitting according to a power-law distribution. The ideal intensity profiles are plotted in solid lines, and the convolved intensity profiles are plotted in dashed lines. The viewing angle is $15^{\circ}$. The values of $\tau_{\rm max} < 1$ are plotted in the right y-axis.}
\label{fig:s200}
\end{figure}

In Figure \ref{fig:blazars_radio_tau1}, we summarize the convolved intensity profiles at $\tau_{\rm max} = 1$ for the models with all jet plasma emitting according to a power-law distribution. The upper plot describes the viewing angle of $5^{\circ}$, and the lower plot describes the viewing angle of $15^{\circ}$. We chose $\tau_{\rm max} = 1$ to reflect the intensity behaviour close to radio core, though stratified models cannot be directly related to the Blandford-K{\"o}nigl model. The models with small $d_{R_{\rm L}}$ values (low BH spins) and high initial magnetization $\sigma_{\rm M}$ provide the asymmetries, and no model demonstrates any distinguishable symmetric limb brightening. The best candidate for self-consistent limb brightening is the model with the total current closed inside a jet at high magnetization and high BH spin (see lower right part in Figure \ref{fig:radio_spin}). In Figure \ref{fig:s200}, we additionally considered $\sigma_{\rm M} = 200$ to demonstrate that the combination of higher magnetization and BH spin indeed results qualitatively in more symmetrical distinguishable limb brightening.

Regarding the magnetization and BH spin predictions, similarly to our work, \citet{BroderickLoeb2009} showed that low BH spins lead to strong asymmetries and high BH spins can produce limb brightening; \citet{Fuentesetal2018} showed triple-peaked intensity profiles in high magnetized jets. Helical magnetic fields in RMHD simulations by \citet{KramerMacDonald2021} produced asymmetric intensity maps, while the magnetic fields in the models we considered also contain both poloidal and toroidal components. 

Nevertheless, the profiles with qualitatively notable transverse structures are still optically thick and excessively bright. This questions the explanation of the multipeaked intensity profiles solely by the high Lorentz-factors. The following reasoning also takes place. 
Within the Blandford-K{\"o}nigl (BK) model \citep{BlandfordKoenigl1979}, the magnetic field and the amplitude of the particle number density at distance $r$ are given by the relations
\begin{equation}
\label{eq:BK_B}
    B(r) = B_0 \left(\frac{r_0}{r}\right),
\end{equation}
\begin{equation}
\label{eq:BK_ke}
    k_e(r) = k_{e, 0}(r) \left(\frac{r_0}{r}\right)^2.
\end{equation}

These dependencies assume the homogeneity of the source and the equipartition between the magnetic and emitting particles energy densities. Though we consider the stratified models, we can compare their predictions with the observations using the BK model. We set $1$ G and $10^3$ cm$^{-3}$ as fiducial values for the magnetic field and particle number density at 1 pc \citep[e.g.][]{L98}. To compare, we average the magnetic field and emitting particle number density for our $15^{\circ}$ models at $\tau_{\rm max} = 1$ and calculate the emitting particle number density predicted by BK model for such magnetic fields (last two lines in Tables \ref{tab:radio_spin}, \ref{tab:radio_BHspin_l0.01l0.95}, \ref{tab:radio_BHspin_Ohmic}). For all plasma emitting (Table \ref{tab:radio_spin}), the means of the number density maxima, $\langle n'_{\rm max} \rangle$, and of the averaged per-model number densities, $\langle n'_{\rm average} \rangle$, are two and one order of magnitude higher than predicted correspondingly. We keep in mind that the particle number density predicted in BK by the averaged magnetic field would be even lower. This suggests that not all the plasma in the jet emits. For 1 per cent of the emitting particles (Table \ref{tab:radio_BHspin_l0.01l0.95}), $\langle n'_{\rm max} \rangle$ is one order of magnitude higher than predicted by BK model, and $\langle n'_{\rm average} \rangle$ has the same order of magnitude as predicted. So, the suppressed region of the jet should indeed occupy almost whole cross-section. For Ohmic heating (Table \ref{tab:radio_BHspin_Ohmic}), $\langle n'_{\rm max} \rangle$ is very high as the proper electric current tends to delta-like behaviour at high magnetization and BH spin \autoref{ss:Ohm}, so  $\langle n'_{\rm average} \rangle$ is more applicable for comparison. This value differs from the predicted by BK model by less than two times. Thus, among the all plasma emitting, 1 per cent emitting and Ohmic heating, Ohmic heating provides the best correspondence between equipartition-predicted and actual emitting plasma number density.

\begin{figure}
\centering
\includegraphics[width=\columnwidth, angle=0]{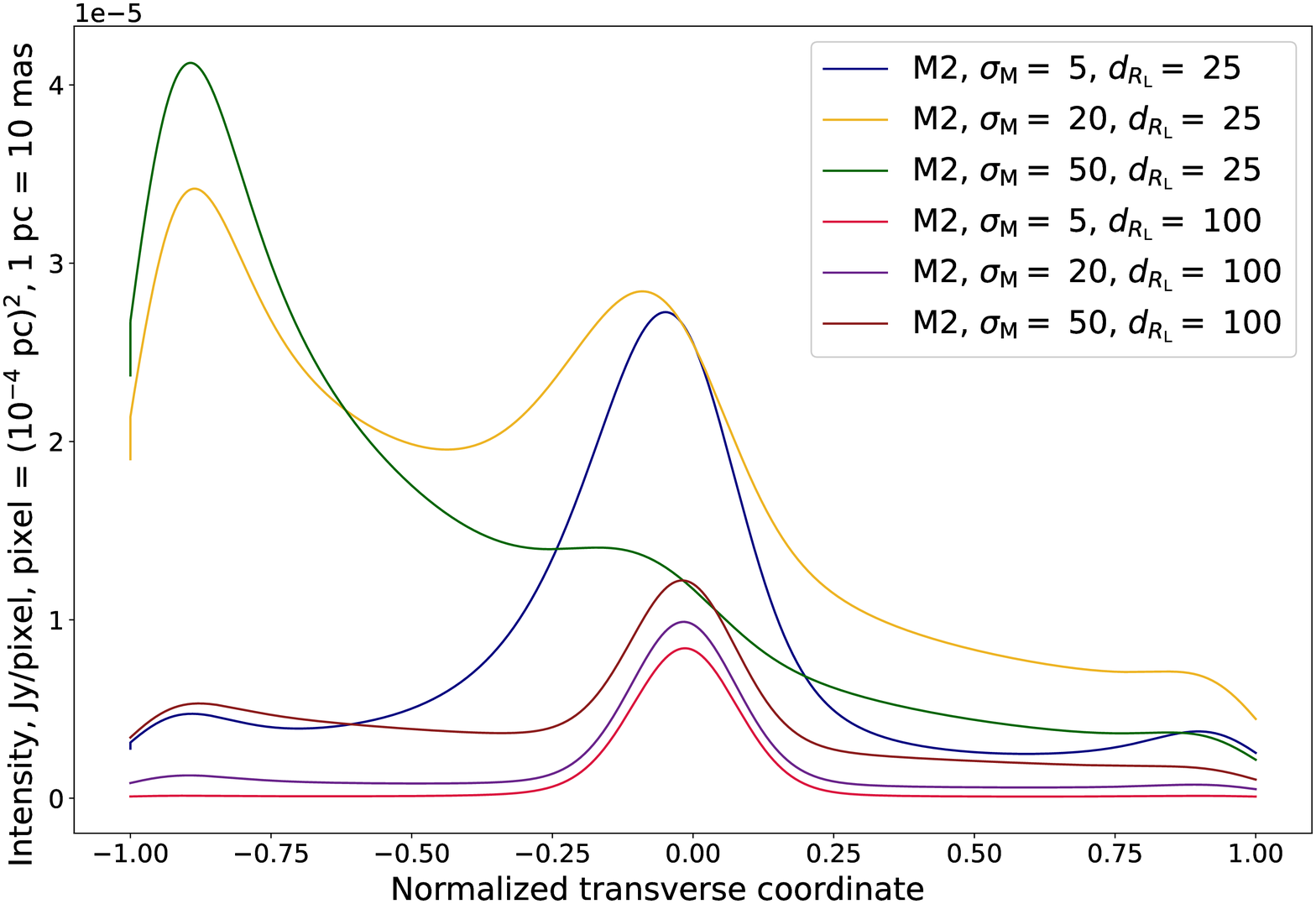}
\includegraphics[width=\columnwidth, angle=0]{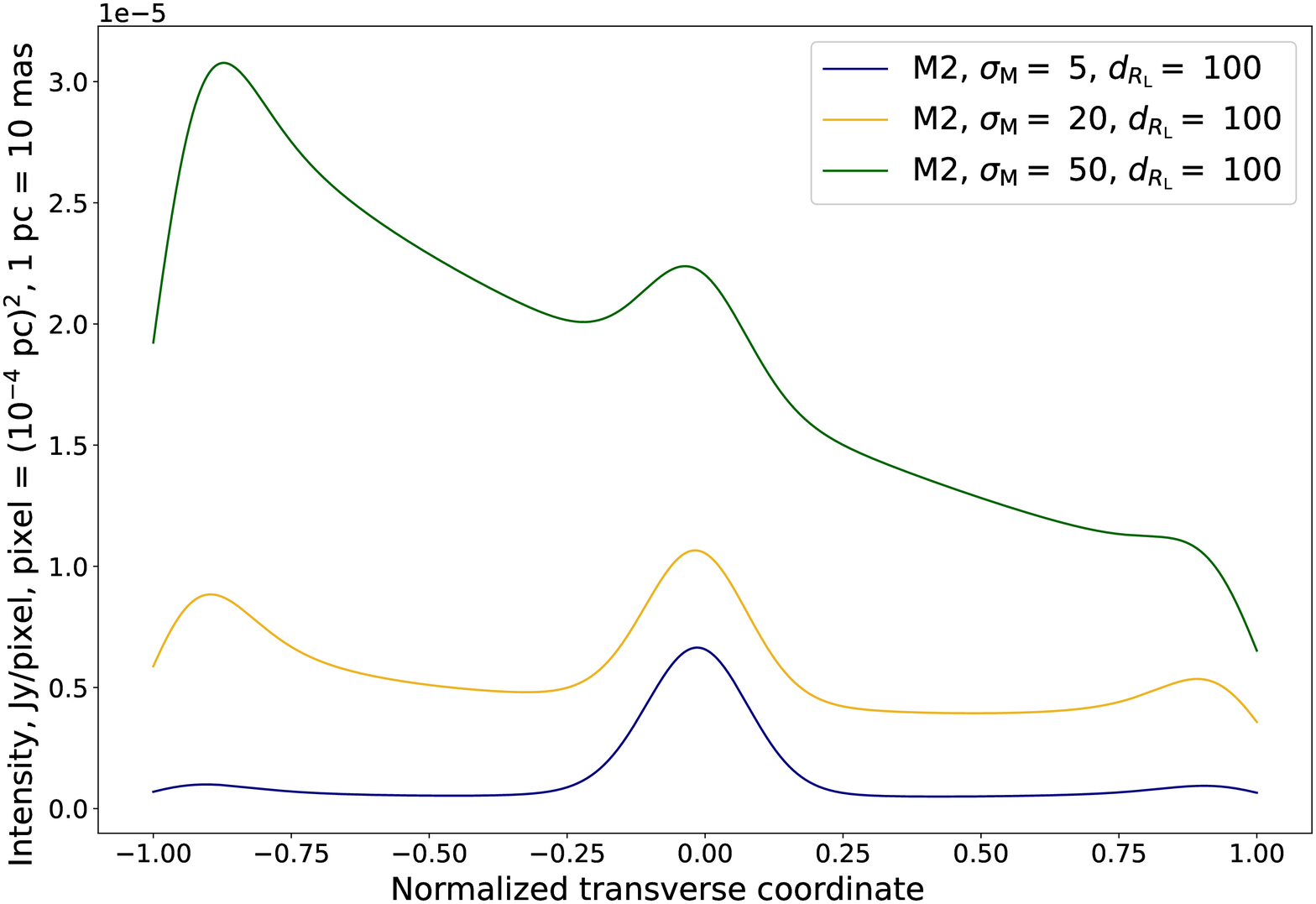}
\caption{The convolved intensity profiles at $\tau_{\rm max} = 1$ for the models with the suppressed central core. In the upper plot, only $1$ per cent of central core plasma emits, and in the lower plot it is only $0.1$ per cent. The viewing angle is $15^{\circ}$.}
\label{fig:l0.01l.95_l0.001l0.95_tau1}
\end{figure}

Regarding the intensity profiles, in Figure \ref{fig:l0.01l.95_l0.001l0.95_tau1}, we show the convolved intensity profiles at $\tau_{\rm max} = 1$ for the models with the suppressed central core. In the upper plot, only $1$ per cent of central core plasma emits, and in the lower plot it is only $0.1$ per cent. The viewing angle is $15^{\circ}$. The profiles demonstrate that a qualitative triple-peaked structure indeed requires this degree of central core suppression. We note that the averaged optical depths for these $\tau_{\rm max} = 1$ sources have the order of $0.01$. The limb-brightened intensity profiles obtained by \citet{Takahashi2018, Ogihara19} also required the enforcing of the emitting plasma transverse stratification.

\begin{figure}
\centering
\includegraphics[width=\columnwidth, angle=0]{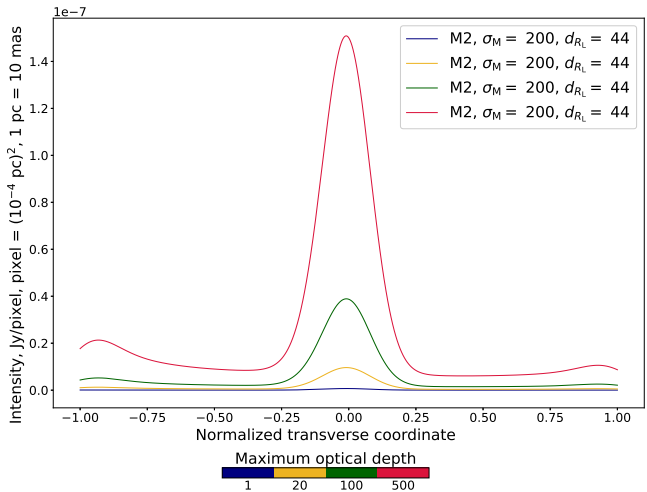}
\caption{The convolved intensity profiles in the high magnetized ($\sigma_{\rm M} = 200$) models where the emitting plasma is distributed according to Ohmic heating. The viewing angle is $15^{\circ}$.}
\label{fig:strong_Ohmic}
\end{figure}

In Figure \ref{fig:strong_Ohmic}, we show the convolved intensity profiles for the high magnetized ($\sigma_{\rm M} = 200$) models where the emitting plasma is distributed according to Ohmic heating. We did not choose the particular $\tau_{\rm max}$ value here, as the Ohmic heated region is extremely narrow. Particularly, at $\tau_{\rm max} = 500$ here, $\tau_{\rm average}$ is just $\approx 6 \cdot 10^{-4}$, so such source is actually optically thin. The plot shows that Ohmic heating provides symmetric, limb-brightened profiles, and the net intensity profile is the most realistic among the cases considered in the paper. 

Other derived parameters for the models are presented in the tables \ref{tab_app:blazars_spin}, \ref{tab_app:radio_spin}, \ref{tab_app:radio_BHspin_l0.01l0.95}, \ref{tab_app:radio_BHspin_Ohmic}, \ref{tab_app:evolution}. In order not to overload the reader with similar data, we limited the tables to three distinctive $\tau_{\rm max}$ values: $0.01$, $1$, $100$. Excluding few of the optically thick cases, we see that the total magnetic flux has the order of $10^{33}-10^{34}$ G cm$^2$ and the jet power mostly has the order of $10^{43}-10^{44}$ erg s$^{-1}$, with few cases of $10^{42}$ erg s$^{-1}$ for a weakly magnetized jets and of $10^{45}$ erg s$^{-1}$ for a strongly magnetized jets.

Thus, we conclude that likely not all of jet plasma is involved in the radiation, though the mechanism should be investigated further. In Frolova et al. in prep., we are deriving the particular fitting parameters for radio galaxy M87 and quasar 3C~273.

\section{Summary and results}
\label{s:summary}

Recent VLBI observations have revealed the AGNs with multi-peaked transverse intensity profiles, among them Mrk501 \citep{Giroletti04}, M87 \citep{Hada17, Walker18}, 3C 84 \citep{Giov18}, Cen A \citep{Janssen21}, 3C 273 \citep{Bruni2021}.  
In this paper, we investigate the possible origins of such intensity profiles by calculating self-absorbed synchrotron emission from the stratified jets. The jet structure is obtained within semi-analitical magnetohydrodynamical models with a constant angular velocity (M1, \citet{Beskin06, Lyu09}) and with the current closed inside the jet (M2, \citet{BCKN-17, Kov-20}). 

We obtained the following results. 
If we assume that all the particles in a jet, obtained within the MHD models, emit, then the jet emission must be dominated by the dense central part of the jet -- the central core. Its existence has been predicted in both analytical \citep{Beskin09} and numerical models \citep{Kom09}, thus, it is a robust result of MHD modelling. For most of the model parameters the central core results in spine-brightened profiles with weak and asymmetrical edge effects, if present. However, the total spectral flux in this case exceeds the typical observed value by several orders of magnitude. Thus, we conclude, that if the nonthermal particles spatial distribution is proportional to the total particle number density, then only a small fraction of all plasma emits. This is in agreement with the models of particle acceleration on shocks or in reconnection events \citep{SironiSpitkovsky2011, SironiSpitkovskyArons2013, SironiSpitkovsky2014}.  

For such spatial distribution of the nonthermal particles, the double-peaked intensity profile appears in optically thick jets, transiting into one-peak profile for more optically thin regime. It can qualitatively explain the spine-to-limb brightening transition at $1.6-4.8$ GHz in 3C 273 observed by \citet{Bruni2021}.

For a jets with large initial magnetization (and large bulk flow Lorentz factor), we observe the shift of intensity peak to the receding part of a jet. In this case we predict that multifrequency polarimetric observations must yield the Faraday rotation map with only one sign, as is observed by \citet{Gabuzda2017, Gabuzda21}. As we move along a jet, the peak shifts closer to the jet axis. This may lead to an apparent change in the jet position angle even for a intrinsically straight jets.  

We observe that in stratified jets, the limb brightening cannot be solely attributed to the presence of a slow flow at the jet boundary. Although the de-boosting of the edge emission provides some increase to the observed intensity, the overall decrease of the particle number density and magnetic field suppresses it. In order to explain the sufficient brightening of the jet edges, one needs low velocity, high particle number density and high magnetic field. If the first two conditions can be fulfilled considering the disk wind, obtaining the necessary high magnetic field may be problematic. Thus, we conclude, that spatial nonthermal particle distribution cannot explain the observed jet emission features.

We explore different spatial distributions of nonthermal particles. We find that the model with distribution (\ref{Ohmic}), suggested by \citet{Lyutikov2005}, is a promising one: i. it provides the values of a spectral flux at the observed levels; ii. the model with a total electric current closed inside a jet yields the triple-peaked intensity profile; iii. the intensity distribution is symmetrical. On the other hand, a model with the emitting particle number density proportional to the magnetic energy density in the plasma proper frame results in a strong central core dominance. Thus, this model is unlikely.

We also consider the following emitting particles spatial distribution: up to some radius some fixed fraction of all particles radiate. This suppresses the central core emission, while leaving such interesting features as a change in a jet position angle and one sign of Faraday Rotation Measure. Outward from this radius all the plasma emits. This allows us to explain the triple-peaked intensity profile and the transition from triple-peaked to single-peaked profiles down the jet. Such spatial distribution can be advocated by the following two reasons. First, if the formation of the nonthermal relativistic plasma distribution is due to the magnetic field dissipation (for example, reconnection), then the plasma in a central core would remain non-relativistic and non-emitting, as it dominates the magnetic field in this region. Second, the particles acceleration may be due to reconnection events as a result of instabilities, and such events are expected near the jet boundary. In this case the nonthermal emitting particles appear mainly at the jet edges.

We conclude that the modelling of the observed intensity profiles in the resolved nearby jets by self-absorbed synchrotron emission from the stratified jets constrains the spatial distribution of the emitting plasma. We suggest that the models described in \autoref{ss:spatial} and \autoref{ss:Ohm} are the preferred ones. In the first case the particles must be accelerated with the formation of a power-law energy distribution mainly at the jet edges due to the interaction with the collimating medium (e.g. instabilities triggering the magnetic reconnection). In the second case, the spatial distribution of the emitting particles forms due to Ohmic heating. The future work on the quantitative reproduction of intensity profiles observed in M87 by \citet{Asada2016} and Nikonov et al. submitted and in 3C 273 \citep{Bruni2021} together with the polarization images modelling will allow us to further constrain the possible mechanisms of plasma heating. 

\section*{Acknowledgements}
We thank the referee, Joana Kramer, for helpful comments and suggestions that significantly improved the manuscript.
This study has been supported
by the Russian Science Foundation: project\footnote{Information about the project: \url{https://rscf.ru/en/project/20-72-10078/}} 20-72-10078.
This research made use of the data from the MOJAVE database\footnote{\url{https://www.cv.nrao.edu/MOJAVE/}} which is maintained by the MOJAVE team \citep{MOJAVE_XV}.
This research made use of NASA's Astrophysics Data System.

\section*{Data Availability}
The data underlying this research will be shared on reasonable request to the corresponding author.

\bibliographystyle{mnras}
\bibliography{nee1}
% Don't change these lines
\bsp    % typesetting comment
\label{lastpage}

\appendix

\section{The optically thin and thick physical parameters}
We present the characteristic physical parameters in substantially optically thin ($\tau_{\rm max} = 0.01$) and thick ($\tau_{\rm max} = 100$) conditions for Figures \ref{fig:blazars_spin} (Table \ref{tab_app:blazars_spin}), \ref{fig:radio_spin} (Table \ref{tab_app:radio_spin}), \ref{fig:radio_BHspin_l0.01l0.95} (Table \ref{tab_app:radio_BHspin_l0.01l0.95}), \ref{fig:radio_BHspin_Ohmic} (Table \ref{tab_app:radio_BHspin_Ohmic}) and for Figure \ref{fig:evolution} representing the jet at different distances from the central engine (Table \ref{tab_app:evolution}).

\begin{table*}
	\centering
	\begin{tabular}{c|c|c|c|c|c|c|c|c|c|c}
\hline 
		 M & $\sigma_{\rm M}$ & $d_{R_{\rm L}}$ & $\tau_{\rm max}$ & \begin{tabular}[x]{@{}c@{}}$\Psi_0$, \\ $10^{34}$ G cm$^2$\end{tabular} & \begin{tabular}[x]{@{}c@{}}$W_{\rm jet}$, \\ $10^{44}$ erg s$^{-1}$ \end{tabular} & \begin{tabular}[x]{@{}c@{}} $n_{\rm max}^{\prime}$, \\ cm$^{-3}$\end{tabular} & \begin{tabular}[x]{@{}c@{}} $n_{\rm average}^{\prime}$, \\ cm$^{-3}$\end{tabular} & \begin{tabular}[x]{@{}c@{}} $B_{\rm max}^{\prime}$, \\ G\end{tabular}  & \begin{tabular}[x]{@{}c@{}} $B_{\rm average}^{\prime}$,\\ G\end{tabular}  &  \begin{tabular}[x]{@{}c@{}}$I_{\rm max}$, \\ $\mu$Jy/pixel \end{tabular}\\ \hline 
		M1 & 5 & 25 & 0.01 & 0.45 & 0.08 & 145 & 16 & 0.017 & 0.004 & 1.90\\ 
		M1 & 5 & 25 & 100 & 4.47 & 7.98 & 14529 & 1557 & 0.173 & 0.044 & 216.00\\ 
		M1 & 20 & 25 & 0.01 & 0.79 & 0.25 & 81 & 9 & 0.022 & 0.006 & 2.51\\ 
		M1 & 20 & 25 & 100 & 7.94 & 25.15 & 8069 & 915 & 0.217 & 0.063 & 326.86\\ 
		M1 & 50 & 25 & 0.01 & 1.35 & 0.72 & 60 & 7 & 0.024 & 0.008 & 3.83\\ 
		M1 & 50 & 25 & 100 & 13.48 & 72.49 & 5995 & 713 & 0.237 & 0.079 & 323.53\\ 
		M1 & 5 & 100 & 0.01 & 0.06 & 0.02 & 372 & 10 & 0.022 & 0.002 & 0.65\\ 
		M1 & 5 & 100 & 100 & 0.56 & 2.03 & 37168 & 1024 & 0.219 & 0.019 & 163.44\\ 
		M1 & 20 & 100 & 0.01 & 0.11 & 0.08 & 224 & 6 & 0.027 & 0.003 & 0.88\\ 
		M1 & 20 & 100 & 100 & 1.10 & 7.67 & 22368 & 640 & 0.272 & 0.028 & 249.28\\ 
		M1 & 50 & 100 & 0.01 & 0.20 & 0.27 & 181 & 5 & 0.029 & 0.003 & 1.17\\ 
		M1 & 50 & 100 & 100 & 2.05 & 26.81 & 18063 & 537 & 0.294 & 0.034 & 293.00\\ 
		M2 & 5 & 25 & 0.01 & 0.27 & 0.03 & 83 & 11 & 0.023 & 0.006 & 1.96\\ 
		M2 & 5 & 25 & 100 & 2.67 & 2.84 & 8285 & 1095 & 0.225 & 0.063 & 174.45\\ 
		M2 & 20 & 25 & 0.01 & 0.77 & 0.24 & 51 & 7 & 0.026 & 0.009 & 4.29\\ 
		M2 & 20 & 25 & 100 & 7.73 & 23.84 & 5071 & 714 & 0.259 & 0.092 & 195.24\\ 
		M2 & 50 & 25 & 0.01 & 1.15 & 0.53 & 28 & 4 & 0.025 & 0.010 & 5.79\\ 
		M2 & 50 & 25 & 100 & 11.54 & 53.16 & 2763 & 404 & 0.245 & 0.099 & 177.24\\ 
		M2 & 5 & 100 & 0.01 & 0.04 & 0.01 & 219 & 7 & 0.028 & 0.003 & 0.76\\ 
		M2 & 5 & 100 & 100 & 0.36 & 0.81 & 21948 & 687 & 0.279 & 0.028 & 184.73\\ 
		M2 & 20 & 100 & 0.01 & 0.12 & 0.09 & 153 & 5 & 0.031 & 0.004 & 1.50\\ 
		M2 & 20 & 100 & 100 & 1.21 & 9.33 & 15300 & 499 & 0.312 & 0.039 & 323.09\\ 
		M2 & 50 & 100 & 0.01 & 0.25 & 0.41 & 129 & 4 & 0.033 & 0.005 & 2.22\\ 
		M2 & 50 & 100 & 100 & 2.53 & 40.95 & 12877 & 437 & 0.325 & 0.049 & 321.51\\ 
 \hline 
	\end{tabular}
\caption{The physical parameters in substantially optically thin ($\tau_{\rm max} = 0.01$) and thick ($\tau_{\rm max} = 100$) conditions in Figure \ref{fig:blazars_spin}.}
 \label{tab_app:blazars_spin}
\end{table*}

\begin{table*}
	\centering
	\begin{tabular}{c|c|c|c|c|c|c|c|c|c|c}
 \hline
		 M & $\sigma_{\rm M}$ & $d_{R_{\rm L}}$ & $\tau_{\rm max}$ & \begin{tabular}[x]{@{}c@{}}$\Psi_0$, \\ $10^{34}$ G cm$^2$\end{tabular} & \begin{tabular}[x]{@{}c@{}}$W_{\rm jet}$, \\ $10^{44}$ erg s$^{-1}$ \end{tabular} & \begin{tabular}[x]{@{}c@{}} $n_{\rm max}^{\prime}$, \\ cm$^{-3}$\end{tabular} & \begin{tabular}[x]{@{}c@{}} $n_{\rm average}^{\prime}$, \\ cm$^{-3}$\end{tabular} & \begin{tabular}[x]{@{}c@{}} $B_{\rm max}^{\prime}$, \\ G\end{tabular}  & \begin{tabular}[x]{@{}c@{}} $B_{\rm average}^{\prime}$,\\ G\end{tabular}  &  \begin{tabular}[x]{@{}c@{}}$I_{\rm max}$, \\ $\mu$Jy/pixel \end{tabular}\\ \hline 
		M1 & 5 & 25 & 0.01 & 0.56 & 0.13 & 228 & 24 & 0.022 & 0.005 & 1.07\\ 
		M1 & 5 & 25 & 100 & 5.60 & 12.50 & 22755 & 2439 & 0.217 & 0.055 & 129.97\\ 
		M1 & 20 & 25 & 0.01 & 1.02 & 0.41 & 132 & 15 & 0.028 & 0.008 & 1.11\\ 
		M1 & 20 & 25 & 100 & 10.16 & 41.20 & 13217 & 1498 & 0.277 & 0.081 & 122.46\\ 
		M1 & 50 & 25 & 0.01 & 1.77 & 1.24 & 103 & 12 & 0.031 & 0.010 & 1.18\\ 
		M1 & 50 & 25 & 100 & 17.65 & 124.30 & 10279 & 1222 & 0.311 & 0.103 & 106.40\\ 
		M1 & 5 & 100 & 0.01 & 0.07 & 0.03 & 578 & 16 & 0.027 & 0.002 & 0.31\\ 
		M1 & 5 & 100 & 100 & 0.70 & 3.16 & 57810 & 1593 & 0.274 & 0.024 & 87.31\\ 
		M1 & 20 & 100 & 0.01 & 0.16 & 0.15 & 449 & 13 & 0.039 & 0.004 & 0.32\\ 
		M1 & 20 & 100 & 100 & 1.55 & 15.38 & 44851 & 1283 & 0.385 & 0.039 & 87.79\\ 
		M1 & 50 & 100 & 0.01 & 0.26 & 0.44 & 293 & 9 & 0.037 & 0.004 & 0.34\\ 
		M1 & 50 & 100 & 100 & 2.61 & 43.55 & 29345 & 872 & 0.374 & 0.043 & 85.10\\ 
		M2 & 5 & 25 & 0.01 & 0.33 & 0.04 & 130 & 17 & 0.028 & 0.008 & 1.11\\ 
		M2 & 5 & 25 & 100 & 3.34 & 4.45 & 12976 & 1716 & 0.282 & 0.079 & 127.83\\ 
		M2 & 20 & 25 & 0.01 & 1.00 & 0.40 & 85 & 12 & 0.034 & 0.012 & 1.33\\ 
		M2 & 20 & 25 & 100 & 10.04 & 40.20 & 8549 & 1203 & 0.336 & 0.119 & 122.12\\ 
		M2 & 50 & 25 & 0.01 & 1.82 & 1.32 & 68 & 10 & 0.039 & 0.016 & 1.48\\ 
		M2 & 50 & 25 & 100 & 18.17 & 131.79 & 6849 & 1002 & 0.386 & 0.157 & 92.13\\ 
		M2 & 5 & 100 & 0.01 & 0.04 & 0.01 & 344 & 11 & 0.035 & 0.003 & 0.33\\ 
		M2 & 5 & 100 & 100 & 0.45 & 1.27 & 34438 & 1077 & 0.350 & 0.035 & 94.83\\ 
		M2 & 20 & 100 & 0.01 & 0.16 & 0.15 & 252 & 8 & 0.040 & 0.005 & 0.37\\ 
		M2 & 20 & 100 & 100 & 1.55 & 15.38 & 25235 & 823 & 0.401 & 0.050 & 88.54\\ 
		M2 & 50 & 100 & 0.01 & 0.33 & 0.70 & 220 & 7 & 0.043 & 0.006 & 0.39\\ 
		M2 & 50 & 100 & 100 & 3.31 & 70.02 & 22019 & 748 & 0.426 & 0.064 & 80.35\\ 
 \hline 
	\end{tabular}
 \caption{The physical parameters in substantially optically thin ($\tau_{\rm max} = 0.01$) and thick ($\tau_{\rm max} = 100$) conditions in Figure \ref{fig:radio_spin}.}
  \label{tab_app:radio_spin}
\end{table*}

\begin{table*}
	\centering
	\begin{tabular}{c|c|c|c|c|c|c|c|c|c|c}
 \hline
		 M & $\sigma_{\rm M}$ & $d_{R_{\rm L}}$ & $\tau_{\rm max}$ & \begin{tabular}[x]{@{}c@{}}$\Psi_0$, \\ $10^{34}$ G cm$^2$\end{tabular} & \begin{tabular}[x]{@{}c@{}}$W_{\rm jet}$, \\ $10^{44}$ erg s$^{-1}$ \end{tabular} & \begin{tabular}[x]{@{}c@{}} $n_{\rm max}^{\prime}$, \\ cm$^{-3}$\end{tabular} & \begin{tabular}[x]{@{}c@{}} $n_{\rm average}^{\prime}$, \\ cm$^{-3}$\end{tabular} & \begin{tabular}[x]{@{}c@{}} $B_{\rm max}^{\prime}$, \\ G\end{tabular}  & \begin{tabular}[x]{@{}c@{}} $B_{\rm average}^{\prime}$,\\ G\end{tabular}  &  \begin{tabular}[x]{@{}c@{}}$I_{\rm max}$, \\ $\mu$Jy/pixel \end{tabular}\\ \hline 
		M2 & 5 & 25 & 0.01 & 1.05 & 0.44 & 151 & 7 & 0.089 & 0.025 & 0.65\\ 
		M2 & 5 & 25 & 100 & 10.51 & 44.03 & 15070 & 674 & 0.887 & 0.250 & 60.52\\ 
		M2 & 20 & 25 & 0.01 & 2.68 & 2.87 & 170 & 5 & 0.090 & 0.032 & 0.79\\ 
		M2 & 20 & 25 & 100 & 26.84 & 287.46 & 16983 & 457 & 0.898 & 0.319 & 50.08\\ 
		M2 & 50 & 25 & 0.01 & 3.18 & 4.04 & 92 & 2 & 0.068 & 0.027 & 0.96\\ 
		M2 & 50 & 25 & 100 & 31.82 & 403.93 & 9191 & 188 & 0.676 & 0.274 & 60.89\\ 
		M2 & 5 & 100 & 0.01 & 0.14 & 0.13 & 44 & 2 & 0.111 & 0.011 & 0.19\\ 
		M2 & 5 & 100 & 100 & 1.41 & 12.69 & 4359 & 219 & 1.107 & 0.110 & 54.42\\ 
		M2 & 20 & 100 & 0.01 & 0.49 & 1.54 & 91 & 2 & 0.127 & 0.016 & 0.22\\ 
		M2 & 20 & 100 & 100 & 4.90 & 153.55 & 9147 & 207 & 1.266 & 0.159 & 59.13\\ 
		M2 & 50 & 100 & 0.01 & 1.04 & 6.94 & 158 & 2 & 0.134 & 0.020 & 0.26\\ 
		M2 & 50 & 100 & 100 & 10.43 & 694.06 & 15810 & 220 & 1.340 & 0.201 & 91.10\\ 

 \hline 
	\end{tabular}
 \caption{The physical parameters in substantially optically thin ($\tau_{\rm max} = 0.01$) and thick ($\tau_{\rm max} = 100$) conditions in Figure \ref{fig:radio_BHspin_l0.01l0.95}.}
 \label{tab_app:radio_BHspin_l0.01l0.95}
\end{table*}

\begin{table*}
	\centering
	\begin{tabular}{c|c|c|c|c|c|c|c|c|c|c}
 \hline
		 M & $\sigma_{\rm M}$ & $d_{R_{\rm L}}$ & $\tau_{\rm max}$ & \begin{tabular}[x]{@{}c@{}}$\Psi_0$, \\ $10^{34}$ G cm$^2$\end{tabular} & \begin{tabular}[x]{@{}c@{}}$W_{\rm jet}$, \\ $10^{44}$ erg s$^{-1}$ \end{tabular} & \begin{tabular}[x]{@{}c@{}} $n_{\rm max}^{\prime}$, \\ cm$^{-3}$\end{tabular} & \begin{tabular}[x]{@{}c@{}} $n_{\rm average}^{\prime}$, \\ cm$^{-3}$\end{tabular} & \begin{tabular}[x]{@{}c@{}} $B_{\rm max}^{\prime}$, \\ G\end{tabular}  & \begin{tabular}[x]{@{}c@{}} $B_{\rm average}^{\prime}$,\\ G\end{tabular}  &  \begin{tabular}[x]{@{}c@{}}$I_{\rm max}$, \\ $\mu$Jy/pixel \end{tabular}\\ \hline 
		M2 & 5 & 25 & 0.01 & 0.42 & 0.07 & 205 & 6 & 0.035 & 0.010 & 9.58\\ 
		M2 & 5 & 25 & 100 & 4.19 & 7.02 & 20480 & 646 & 0.354 & 0.100 & 54.94\\ 
		M2 & 20 & 25 & 0.01 & 1.42 & 0.80 & 171 & 4 & 0.047 & 0.017 & 3.20\\ 
		M2 & 20 & 25 & 100 & 14.19 & 80.32 & 17083 & 370 & 0.475 & 0.169 & 54.47\\ 
		M2 & 50 & 25 & 0.01 & 3.12 & 3.88 & 202 & 2 & 0.066 & 0.027 & 0.26\\ 
		M2 & 50 & 25 & 100 & 31.18 & 387.80 & 20156 & 213 & 0.662 & 0.269 & 48.02\\ 
		M2 & 5 & 100 & 0.01 & 0.05 & 0.02 & 505 & 4 & 0.042 & 0.004 & 0.11\\ 
		M2 & 5 & 100 & 100 & 0.54 & 1.86 & 50540 & 445 & 0.424 & 0.042 & 17.48\\ 
		M2 & 20 & 100 & 0.01 & 0.20 & 0.25 & 403 & 3 & 0.051 & 0.006 & 0.10\\ 
		M2 & 20 & 100 & 100 & 1.96 & 24.55 & 40286 & 313 & 0.506 & 0.064 & 18.72\\ 
		M2 & 50 & 100 & 0.01 & 0.43 & 1.18 & 372 & 3 & 0.055 & 0.008 & 0.10\\ 
		M2 & 50 & 100 & 100 & 4.30 & 118.15 & 37155 & 264 & 0.553 & 0.083 & 20.11\\ 

 \hline 
	\end{tabular}
 \caption{The physical parameters in substantially optically thin ($\tau_{\rm max} = 0.01$) and thick ($\tau_{\rm max} = 100$) conditions in Figure \ref{fig:radio_BHspin_Ohmic}.}
 \label{tab_app:radio_BHspin_Ohmic}
\end{table*}

\section{The physical parameters at different distances from the central engine}
We present the characteristic physical parameters for the maximum optical depth of $\tau_{\rm max} = 1$ and in substantially optically thin ($\tau_{\rm max} = 0.01$) and thick ($\tau_{\rm max} = 100$) conditions for Figure \ref{fig:evolution} representing the jet at different distances from the central engine (Table \ref{tab_app:evolution}).

\begin{table*}
	\centering
	\begin{tabular}{c|c|c|c|c|c|c|c|c|c|c}
 \hline
		 M & $\sigma_{\rm M}$ & $d_{R_{\rm L}}$ & $\tau_{\rm max}$ & \begin{tabular}[x]{@{}c@{}}$\Psi_0$, \\ $10^{34}$ G cm$^2$\end{tabular} & \begin{tabular}[x]{@{}c@{}}$W_{\rm jet}$, \\ $10^{44}$ erg s$^{-1}$ \end{tabular} & \begin{tabular}[x]{@{}c@{}} $n_{\rm max}^{\prime}$, \\ cm$^{-3}$\end{tabular} & \begin{tabular}[x]{@{}c@{}} $n_{\rm average}^{\prime}$, \\ cm$^{-3}$\end{tabular} & \begin{tabular}[x]{@{}c@{}} $B_{\rm max}^{\prime}$, \\ G\end{tabular}  & \begin{tabular}[x]{@{}c@{}} $B_{\rm average}^{\prime}$,\\ G\end{tabular}  &  \begin{tabular}[x]{@{}c@{}}$I_{\rm max}$, \\ $\mu$Jy/pixel \end{tabular}\\ \hline 
		M1 & 5 & 50 & 0.01 & 0.45 & 0.08 & 112 & 6 & 0.013 & 0.002 & 0.94\\ 
		M1 & 5 & 50 & 1 & 1.41 & 0.80 & 1120 & 61 & 0.042 & 0.006 & 46.00\\ 
		M1 & 5 & 50 & 100 & 4.47 & 7.98 & 11196 & 612 & 0.133 & 0.020 & 194.21\\ 
		M1 & 5 & 100 & 0.01 & 0.45 & 0.08 & 91 & 3 & 0.011 & 0.001 & 0.54\\ 
		M1 & 5 & 100 & 1 & 1.41 & 0.80 & 912 & 25 & 0.034 & 0.003 & 28.35\\ 
		M1 & 5 & 100 & 100 & 4.47 & 7.98 & 9119 & 251 & 0.109 & 0.010 & 177.07\\ 
		M1 & 50 & 50 & 0.01 & 1.35 & 0.72 & 40 & 2 & 0.016 & 0.003 & 0.89\\ 
		M1 & 50 & 50 & 1 & 4.26 & 7.25 & 402 & 24 & 0.050 & 0.010 & 45.66\\ 
		M1 & 50 & 50 & 100 & 13.48 & 72.49 & 4017 & 239 & 0.159 & 0.032 & 357.26\\ 
		M1 & 50 & 100 & 0.01 & 1.35 & 0.72 & 31 & 1 & 0.012 & 0.001 & 0.36\\ 
		M1 & 50 & 100 & 1 & 4.26 & 7.25 & 305 & 9 & 0.038 & 0.004 & 19.53\\ 
		M1 & 50 & 100 & 100 & 13.48 & 72.49 & 3052 & 91 & 0.121 & 0.014 & 259.04\\ 
		M2 & 50 & 50 & 0.01 & 1.15 & 0.53 & 15 & 1 & 0.013 & 0.004 & 0.71\\ 
		M2 & 50 & 50 & 1 & 3.65 & 5.32 & 151 & 11 & 0.042 & 0.011 & 38.44\\ 
		M2 & 50 & 50 & 100 & 11.54 & 53.16 & 1513 & 109 & 0.134 & 0.035 & 374.31\\ 
		M2 & 50 & 100 & 0.01 & 1.15 & 0.53 & 10 & 0 & 0.009 & 0.001 & 0.14\\ 
		M2 & 50 & 100 & 1 & 3.65 & 5.32 & 104 & 4 & 0.029 & 0.004 & 7.87\\ 
		M2 & 50 & 100 & 100 & 11.54 & 53.16 & 1045 & 35 & 0.093 & 0.014 & 245.54\\ 
		M1 & 50 & 50 & 0.01 & 1.77 & 1.24 & 69 & 4 & 0.021 & 0.004 & 0.39\\ 
		M1 & 50 & 50 & 1 & 5.58 & 12.43 & 689 & 41 & 0.066 & 0.013 & 19.83\\ 
		M1 & 50 & 50 & 100 & 17.65 & 124.30 & 6888 & 410 & 0.208 & 0.041 & 115.86\\ 
		M1 & 50 & 100 & 0.01 & 1.77 & 1.24 & 52 & 2 & 0.016 & 0.002 & 0.18\\ 
		M1 & 50 & 100 & 1 & 5.58 & 12.43 & 523 & 16 & 0.050 & 0.006 & 9.83\\ 
		M1 & 50 & 100 & 100 & 17.65 & 124.30 & 5234 & 156 & 0.158 & 0.018 & 107.68\\ 
		M2 & 50 & 50 & 0.01 & 1.82 & 1.32 & 38 & 3 & 0.021 & 0.006 & 0.27\\ 
		M2 & 50 & 50 & 1 & 5.75 & 13.18 & 375 & 27 & 0.067 & 0.018 & 14.36\\ 
		M2 & 50 & 50 & 100 & 18.17 & 131.79 & 3751 & 270 & 0.211 & 0.055 & 109.39\\ 
		M2 & 50 & 100 & 0.01 & 1.82 & 1.32 & 26 & 1 & 0.015 & 0.002 & 0.09\\ 
		M2 & 50 & 100 & 1 & 5.75 & 13.18 & 259 & 9 & 0.046 & 0.007 & 5.15\\ 
		M2 & 50 & 100 & 100 & 18.17 & 131.79 & 2590 & 88 & 0.146 & 0.022 & 96.31\\ 
 \hline 
	\end{tabular}
 \caption{The parameters for Figure \ref{fig:evolution} (see also Tables \ref{tab:blazars_spin}, \ref{tab:radio_spin}, \ref{tab_app:blazars_spin}, \ref{tab_app:radio_spin}).}
  \label{tab_app:evolution}
\end{table*}

\end{document}